\documentclass[fleqn,usenatbib]{mnras}

\usepackage{newtxtext,newtxmath}
\usepackage{orcidlink}
\usepackage{verbatim}

\usepackage[T1]{fontenc}
\usepackage[utf8]{inputenc}

\DeclareRobustCommand{\VAN}[3]{#2}
\let\VANthebibliography\thebibliography
\def\thebibliography{\DeclareRobustCommand{\VAN}[3]{##3}\VANthebibliography}

\usepackage{graphicx}	
\usepackage{amsmath, mathtools, nccmath, nicefrac}	
\usepackage{gensymb} 
\usepackage{enumitem}
\usepackage{textcomp}
\usepackage{chngpage}
\usepackage{footmisc}
\usepackage{bm}
\usepackage{xspace}
\usepackage{tablefootnote, threeparttable}
\usepackage{subfigure}
\usepackage{pdflscape}
\usepackage{float}
\usepackage{makecell}
\usepackage[makeroom]{cancel}
\usepackage{CJK}
\usepackage{arydshln} 

\PassOptionsToPackage{usenames, dvipsnames}{color}
\usepackage{hyperref}
\definecolor{dblue}{RGB}{40, 116, 166}
\definecolor{navy}{RGB}{56, 70, 184}
\definecolor{darkblue}{RGB}{39, 76, 119}
\definecolor{grey}{RGB}{192,192,192}
\definecolor{new_color}{HTML}{CF0000} 
\definecolor{crimson}{RGB}{220,20,60}

\hypersetup{
    colorlinks = true,
    linkcolor = black,
    citecolor= dblue,
    urlcolor = cyan,
}

\newcommand{\masy}{mas\,y$^{-1}$}
\newcommand{\Msun}{\mbox{$M_{\odot}$}}

\newcommand{\Rearth}{R$_{\oplus}$}
\newcommand{\Mearth}{M$_{\oplus}$}
\newcommand{\cecilia}{\texttt{cecilia}}
\newcommand{\angs}{\text{\normalfont\AA}}
\newcommand{\Gaia}{\emph{Gaia}}
\newcommand{\Galex}{\emph{GALEX}}
\newcommand{\panstarrs}{\emph{Pan-STARRS}}
\newcommand{\wdreal}{WD 1232+563}

\newcommand{\Robs}{$R_{\rm obs}$} 
\newcommand{\wobs}{$\lambda_{\rm obs}$} 
\newcommand{\fobs}{$f_{\rm obs}$}
\newcommand{\ferrobs}{$f_{\rm obs, err}$} 
\newcommand{\rvobs}{$v_{\rm obs}$} 

\newcommand{\Rsynth}{$R_{\rm synth}$} 
\newcommand{\wsynth}{$\lambda_{\rm synth}$} 
\newcommand{\fsynth}{$f_{\rm synth}$} 
\newcommand{\wsynthcorr}{$w_{\rm synth, corr}$} 
\newcommand{\fsynthcorr}{$f_{\rm synth, corr}$}

\newcommand{\Teff}{$T_{\rm eff}$}                 
\newcommand{\logg}{$\log\rm g$}                   
\newcommand{\logHHe}{$\log_{10}\rm{(H/He)}$}      
\newcommand{\logCaHe}{$\log_{10}\rm{(Ca/He)}$}    
\newcommand{\logMgHe}{$\log_{10}\rm{(Mg/He)}$}    
\newcommand{\logFeHe}{$\log_{10}\rm{(Fe/He)}$}    
\newcommand{\logOHe}{$\log_{10}\rm{(O/He)}$}      
\newcommand{\logSiHe}{$\log_{10}\rm{(Si/He)}$}    
\newcommand{\logTiHe}{$\log_{10}\rm{(Ti/He)}$}    
\newcommand{\logBeHe}{$\log_{10}\rm{(Be/He)}$}    
\newcommand{\logCrHe}{$\log_{10}\rm{(Cr/He)}$}    
\newcommand{\logMnHe}{$\log_{10}\rm{(Mn/He)}$}    
\newcommand{\logNiHe}{$\log_{10}\rm{(Ni/He)}$}    

\title[Machine Learning Analysis of Polluted White Dwarfs]{\cecilia: A Machine Learning-Based Pipeline for Measuring Metal Abundances of Helium-rich Polluted White Dwarfs}

\author[Badenas-Agusti et al.]{Mariona Badenas-Agusti,$^{1,2,3\thanks{\textrm{E-mail: mbadenas@mit.edu}}}$\orcidlink{0000-0003-4903-567X}
Javier Viaña,$^{2}$\orcidlink{0000-0002-0563-784X}
Andrew Vanderburg,$^{2}$\orcidlink{0000-0001-7246-5438}
Simon Blouin,$^{4}$\orcidlink{0000-0002-9632-1436}
\newauthor
Patrick Dufour,$^{5}$\orcidlink{0000-0003-4609-4500}
Siyi Xu \begin{CJK*}{UTF8}{gbsn}(许\CJKfamily{bsmi}偲\CJKfamily{gbsn}艺\end{CJK*}),$^{6}$\orcidlink{0000-0002-8808-4282}
Lizhou Sha$^{7}$\orcidlink{0000-0001-5401-8079}
\\
$^{1}$ Department of Earth, Atmospheric and Planetary Sciences, Massachusetts Institute of Technology, Cambridge, MA 02139, USA\\
$^{2}$Department of Physics and Kavli Institute for Astrophysics and Space Research, Massachusetts Institute of Technology, Cambridge, MA 02139, USA \\
$^{3}$ MIT William Asbjornsen Albert Memorial Fellow \\
$^{4}$ Department of Physics and Astronomy, University of Victoria, Victoria, BC V8W 2Y2, Canada \\
$^{5}$ D\'epartement de Physique, Universit\'e de Montr\'eal, Montr\'eal, Qu\'ebec H3C 3J7, Canada \\
$^{6}$ Gemini Observatory/NSF's NOIRLab, 670 N. A'ohoku Place, Hilo, HI, 96720, USA \\
$^{7}$ Department of Astronomy, University of Wisconsin-Madison, 475 N Charter St, Madison, WI 53706, USA \\
}

\date{Accepted 2024 February 06. Received 2024 February 06; in original form 2023 December 19}

\pubyear{2024}

\begin{document}
\label{firstpage}
\pagerange{\pageref{firstpage}--\pageref{lastpage}}
\maketitle

\begin{abstract}

Over the past several decades, conventional spectral analysis techniques of polluted white dwarfs have become powerful tools to learn about the geology and chemistry of extrasolar bodies. Despite their proven capabilities and extensive legacy of scientific discoveries, these techniques are however still limited by their manual, time-intensive, and iterative nature. As a result, they are susceptible to human errors and are difficult to scale up to population-wide studies of metal pollution. This paper seeks to address this problem by presenting \cecilia, the first Machine Learning (ML)-powered spectral modeling code designed to measure the metal abundances of intermediate-temperature (10,000$\leq$\Teff$\leq$20,000~K), Helium-rich polluted white dwarfs. Trained with more than 22,000 randomly drawn atmosphere models and stellar parameters, our pipeline aims to overcome the limitations of classical methods by replacing the generation of synthetic spectra from computationally expensive codes and uniformly spaced model grids, with a fast, automated, and efficient neural-network-based interpolator. More specifically, \cecilia{} combines state-of-the-art atmosphere models, powerful artificial intelligence tools, and robust statistical techniques to rapidly generate synthetic spectra of polluted white dwarfs in high-dimensional space, and enable accurate ($\lesssim$0.1~dex) and simultaneous measurements of 14 stellar parameters ---including 11 elemental abundances--- from real spectroscopic observations. As massively multiplexed astronomical surveys begin scientific operations, \cecilia's performance has the potential to unlock large-scale studies of extrasolar geochemistry and propel the field of white dwarf science into the era of Big Data. In doing so, we aspire to uncover new statistical insights that were previously impractical with traditional white dwarf characterisation techniques.
\end{abstract}

\begin{keywords}
stars: white dwarfs - stars: atmospheres - stars: abundances - techniques: spectroscopic - methods: data analysis - methods: observational
\end{keywords}



\section{Introduction}

In recent years, the field of exoplanets has undergone remarkable growth, revolutionising our understanding of the Universe with the discovery of more than 5500 distant worlds.\footnote{NASA Exoplanet Database (\url{https://exoplanets.nasa.gov/}). Last Accessed: January 2024.} Despite this notable progress, the detection of a potentially habitable Earth-sized planet with an Earth-like interior remains elusive.

In our quest to find exoplanets similar to our own, knowledge of a planet's bulk composition will be crucial to understand its formation and evolution history, geology, climate, and habitability conditions. Traditionally, transit photometry \citep{Charbonneau:2020} and Doppler spectroscopy \citep{Campbell:1988} have been used to measure the radii and masses of exoplanets, respectively. These two parameters, when combined, provide the planetary bulk density, which can then be compared to interior models to infer the planet's most likely composition \citep{Zeng:2016}. Nevertheless, this approach is incomplete because it does not account for the specific elemental abundances of the planet. In addition, for planets smaller than $R_{p}$$\sim$4~\Rearth{}, large uncertainties in radius and mass measurements ($>10\%$) pose challenges in distinguishing between different interior compositions \citep{Zeng:2008}. These problems are further complicated by  degeneracies in theoretical mass-radius diagrams, which can yield various combinations of metal, silicate, water/ice, and gas  for the same density measurement \citep{Dorn:2015, Rogers:2010, Seager:2007}.

\newpage
To date, the most direct ---and only--- way of overcoming these ambiguities and probing the composition of small, terrestrial exoplanets is through spectroscopy of externally polluted White Dwarfs (WDs). These objects are the degenerate stellar cores of low- and intermediate-mass ($\leq10$~\Msun) main-sequence stars. With a typical radius similar to the size of the Earth ($R_{\rm{WD}}\approx 1$~\Mearth), and a mass half of that of the Sun  ($M_{\rm{WD}}\approx0.6$~\Msun), they are extremely dense and compact, which causes their atmospheres to have a stratified chemical structure. In other words, light elements (i.e. H and He) are expected to float in the thin upper layers of their atmospheres, while elements heavier than helium (or ``metals,'' e.g. Ca, Mg, Si, Fe) should sink rapidly towards the core with diffusion timescales much shorter than the cooling age of the white dwarf \citep{Paquette:1986, Koester:2009}. In reality, however, observations suggest that between 25$\%$ and 50$\%$ of single white dwarfs are contaminated with heavy elements in their atmospheres \citep{Koester:2014, Zuckerman:2010, Zuckerman:2003}, likely from the recent or ongoing accretion of disrupted exoplanetary debris from their surviving outer planetary systems \citep{Jura:2003, Zuckerman:2003, Zuckerman:2010, Koester:2014, JuraYoung:2014}.

During the past two decades, high-resolution spectroscopy of these ``polluted'' white dwarfs has offered an excellent opportunity to determine the bulk composition of their accreted material. This technique has enabled the detection of more than 20 heavy elements \citep[e.g. see references in Table 1 of][]{Klein:2021}, revealing that most exoplanetary bodies are dry and rocky \citep{Veras:2021}, with some water-rich exceptions \citep[e.g.][]{Farihi:2011, Farihi:2013, Raddi:2015, Xu:2017, Hoskin:2020}. Despite the significance of these discoveries, conventional analysis techniques of polluted white dwarfs entail time-consuming, manual, and iterative work, which makes them prone to human errors. As a result, these techniques may lead to wrong or sub-optimal results, potentially biasing our understanding of the geology and chemistry of white dwarf pollutants. 
With these challenges in mind, it has  been difficult to transition from individual to population-wide studies of metal pollution. As we move into an era of massive multi-object spectroscopic surveys, when tens of thousands of white dwarfs will be observed and many of them are expected to show multiple heavy elements, conventional analysis techniques will thus struggle to keep up.

In this paper, we address the limitations of ``classical'' white dwarf characterisation methods by leveraging the power of Machine Learning (ML). Recently, multiple ML-based tools have been developed to interpolate and efficiently model spectra with the goal of estimating stellar abundances. Notable examples include \texttt{The Cannon} (aimed at deriving stellar properties based on a large training set of reference stars with well-known abundances; \citealt{Ness:2015}), \texttt{The Payne} (designed to estimate 25 elemental abundances of red giants from low-resolution APOGEE spectra; \citealt{Ting:2019}), \texttt{wdtools} (dedicated to the analysis of H-rich white dwarf spectra with no metal pollution; \citealt{Chandra:2020}) and \texttt{TheHotPayne}, which is an adaptation of \texttt{The Payne} to OBA stars \citep{Xiang:2022}. These ML pipelines have undoubtedly shown promising results, but they have not been designed to fit the complex and diverse spectra of polluted white dwarfs. This paper seeks to bridge this gap by presenting the first ML-based code capable of achieving this goal.

Inspired by the groundbreaking contributions of astrophysicist Cecilia Payne and building upon the ML framework of the \texttt{The Payne}, we have chosen to name our code  ``\cecilia{}'' ---or \textit{\textbf{C}lues of \textbf{E}xoplanet \textbf{C}ompositions \textbf{I}nferred from po\textbf{L}luted Wh\textbf{I}tE dw\textbf{A}rfs}. With \cecilia, we combine state-of-the-art atmosphere models with innovative ML methods to improve the performance of conventional white dwarf analysis techniques and ultimately enable large-scale studies of metal pollution. From a design standpoint, \cecilia's ML architecture contributes to this goal by adhering to three fundamental principles: automation, efficiency, and flexibility. As an automated pipeline, \cecilia{} solves the dependence of current techniques on human-based, time-intensive, and iterative procedures. In terms of efficiency, our code harnesses the speed of ML tools to rapidly estimate elemental abundances of polluted white dwarfs with a retrieval accuracy similar to that of traditional methods. Lastly, \cecilia{} has a flexible ML architecture that can be re-trained at any time with new, better, and/or larger models. This also makes our pipeline a particularly useful feature extraction tool beyond the fields of white dwarf and exoplanetary science. 

In this work, we describe the building blocks of \cecilia, its current performance and limitations, and potential opportunities for improvement. The organisation of the paper is as follows. In Section \ref{sec:fundamentals_ml}, we provide a brief introduction to the the concept of deep neural networks. In Section \ref{sec:data}, we present the stellar parameters and model atmospheres used to train, validate, and test our code. Section \ref{sec:methods} explains \cecilia's ML architecture, our training procedure, and our framework for estimating the elemental abundances of polluted white dwarfs from their spectra. In Section \ref{sec:results}, we test and explore \cecilia's functionality with synthetic and real spectroscopic observations. Section \ref{sec:discussion} discusses the strengths and weaknesses of \cecilia{}, possible future work to enhance its retrieval accuracy, and its relevance in the context of massive datasets from wide-field astronomical surveys. Finally, we offer our conclusions in Section \ref{sec:conclusions}.  

\section{Deep Neural Networks} \label{sec:fundamentals_ml}

Since the 1950s \citep{Turin:1950, McCarthy:1995}, the field of modern Artificial Intelligence (AI) has sought to create intelligent systems capable of emulating human skills, both in terms of human reasoning and behavior. In recent years, the rapid development of technology has transformed this theoretical aspiration into a tangible reality, causing an important shift from model- to data-driven scientific analysis. The creation of faster and more powerful computers, coupled with the availability of extremely large and complex datasets (or ``Big Data''), has only fueled the growth of AI, making neural networks increasingly relevant across a variety of disciplines, including astrophysics \citep{Smith:2023}. 

To date, the task of developing computer programs that can learn, extract, and process meaningful insights from observations is typically referred to as Machine Learning \citep{Samuel:1959}. At the core of many ML algorithms are multilayered, fully connected, feed-forward neural networks (NN). These networks are learning systems where information flows \textit{forward} through several layers of artificial nodes called ``neurons.'' A basic NN consists of an input layer, one or more hidden layers, and an output layer. The higher the number of hidden layers (or ``depth''), the more complex the ML architecture becomes \citep{LeCun:2015}. This is the case of \cecilia, which consists of three neural networks with 5-6 layers each. In general, deeper architectures are more capable of recognising more intricate relationships in the data \citep{Goodfellow:2016}.

\begin{figure}
    \centering
      \includegraphics[width=1\columnwidth]{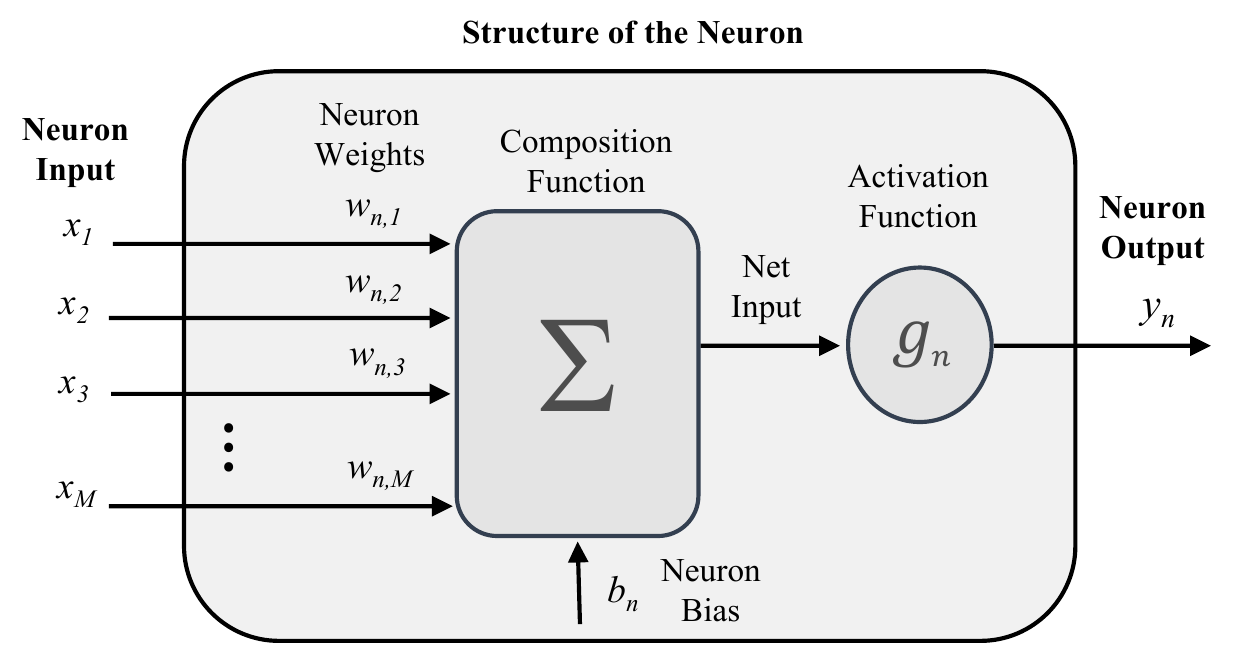}
      \caption{Parameters and operations associated to a single ``neuron,'' which represents the smallest unit of a neural network. }
      \label{fig:neuron}
\end{figure}

In addition to the choice of ML architecture, the learning capabilities of a NN are also shaped by the way its neurons are interconnected with one another. This connection is regulated by three parameters known as ``weights,'' ``biases,'' and ``activation functions.'' The weights and biases determine the importance of the neuronal connections, while the activation functions employ non-linear operations to transform the input data of a set of neurons into useful output parameters. In particular, for a generic neuron $n$, the activation function $g_{n}$ performs the transformation

\begin{equation}
    y_{n} = g_{n}( b_{n} + \sum_{m=1}^{M} w_{n,m} \cdot  x_{m} ),
\end{equation}
where $M$ denotes the total number of inputs fed into the neuron, $m$ is a generic input index, $x$ and $y$ are the input and the output data, and $w_{n,m}$ and $b_{n}$ represent the optimised weight and bias of the neuron, respectively (see \autoref{fig:neuron}). Typically, the activation function $g_{n}$ is chosen by the user experimentally (e.g. see Section \ref{sec:MLarchitecture}).

In the field of ML, the \textit{training} of a NN is an iterative procedure designed to fine-tune the weights and biases of the neurons with the goal of minimising the error (also known as ``loss'' or ``cost'' function) between the network's final predictions and an assumed ground truth. During this process, the NN is exposed to labeled observations in order to identify hidden patterns in the data and learn to generate predictions that closely resemble the expected outcome for a given set of input parameters. As the training proceeds, the predictive accuracy of the NN is  gradually improved with a technique known as ``backpropagation.'' This technique calculates the loss function at each iteration to adjust the model parameters of the network. A standard formulation of the loss function is the quadratic expression
\begin{equation}
    \mathcal{J}_{q} = \frac{1}{2} \sum_{p=1}^{P} (y_{q,p}-\hat{y}_{q,p})^{2} = \frac{1}{2} \sum_{p=1}^{P}e_{q,p}^{2},
\end{equation}
where $q$ is the index of a generic instance,\footnote{In this work, we define an ``instance'' as a spectrum with its corresponding 13 stellar labels.} $P$ is the total number of output neurons,  $p$ is the index of a generic output neuron, $\hat{y}_{q,p}$ is the prediction of the $p$-th output neuron, $y_{q,p}$ is the known true output, and $e$ denotes the residual error difference between $\hat{y}$ and $y$. For a given training dataset, the aggregated loss function of all the instances $Q$ can then be expressed as
\begin{equation}
    \mathcal{J} = \sum_{q=1}^{Q} \mathcal{J}_{q}.
\end{equation}

During training, the aggregated loss function is computed at every step in order to optimise the weights and biases of the network. These parameters are updated iteratively through the gradient of the loss function, $\nabla\mathcal{J}(w_{n,m})$. More specifically, the incremental change of a generic weight in the network, or $\Delta w_{n,m}$, is proportional to the derivative of the total loss function with respect to the current value of the weight,
\begin{equation}
    \Delta w_{n,m} = w_{n,m}^{{\rm new}}-w_{n,m}^{{\rm old}}=-\eta\frac{\partial\mathcal{J}}{\partial w_{n,m}}.
    \label{eq:gradient}
\end{equation}
In this expression, the partial derivative $\partial\mathcal{J}/\partial w_{n,m}$ dictates how much the network parameters should be changed in order to reach a minimum of the loss function, with the negative sign indicating that the direction of the ``descent'' should be against the gradient and towards the minimum. The term $\eta$ is the ``learning rate'' of the network, i.e. a user-defined tuning parameter which provides stability to the learning process and controls the step size of the gradient descent. 

Broadly speaking, the partial derivative $\partial \mathcal{J}/\partial w_{n,m}$ in Eq. \ref{eq:gradient} is proportional to the inputs of the neurons, the derivatives of their activation functions with respect to their weights, and the errors of the network. For the output layer, these factors are decoupled, which makes the computation of the derivative relatively straightforward. However, for the earlier (and deeper) layers, the term $\partial \mathcal{J}/\partial w_{n,m}$ depends on the derivative of the loss function for the outer layers, which in turn is proportional to the \textit{product} of the derivatives of all the activation functions with respect to their weights. This product operation is implemented with the chain rule from the output to the initial layer, i.e. \textit{backwards}, hence giving rise to the concept of \textit{backpropagation}. We note that in many ML applications, including \cecilia, the exact value of the gradient at each training iteration is not computed with the full training sample, but approximated with subsets (or ``batches'') of the latter \citep[e.g.][]{Shallue:2018}.

\section{Data} \label{sec:data}

In this section, we motivate the use of a neural-network spectral interpolator, describe the atmosphere models and synthetic stellar properties used to train, validate, and test \cecilia, and discuss the normalisation techniques used to improve the learning and predictive abilites of our code. 

\subsection{Generation of a Training Set: Motivation}\label{sec:training_set_motivation}

Current white dwarf spectral modelling techniques rely on grids of stellar properties (or ``labels'') with uniform step sizes \citep[e.g.][]{Coutu:2019}. Although this approach has worked well over the years, it can be computationally expensive for a large number of free model parameters. For example, using 13 stellar properties ---as in this work (see Section \ref{sec:synth_labels} below)--- and a relatively coarse grid of 10 points per label, conventional interpolators would require at least 10$^{13}$ models to explore the parameter space of their study. Assuming, on average, that each model takes about 1 hour to generate on a single CPU core, it would take on on the order of 10$^{9}$ CPU-years to compute the whole grid. Therefore, classical methods are prohibitively expensive.


To address this challenge, we haved developed \cecilia, a ML generative spectral interpolator capable of sampling from a high-dimensional label space based on \textit{randomly drawn} training parameters, rather than a uniformly spaced grid of model points. With this ML approach, our pipeline only needs about 20,000 synthetic models to predict white dwarf spectra, i.e. about $2\times 10^{-9}\%$ of the models that classical methods would typically require for the same task. 

\subsubsection{Synthetic Stellar Labels} \label{sec:synth_labels}

In this work, we have chosen to characterise He-rich polluted white dwarfs with 13 stellar properties, namely: their effective temperature (\Teff{}) and surface gravity (\logg{}), their hydrogen abundance, and 10 metal abundances for Ca, Mg, Fe, O, Si, Ti, Be, Cr, Mn, Ni (see Table \ref{tab:ranges_labels}).\footnote{In this work, all the abundances are expressed as logarithmic number abundance ratios (in base 10) relative to helium.} In our models, we have also considered 14 additional heavy elements (C, N, Li, Na, Al, P, S, Cl, Ar, K, Sc, V, Co, and Cu). However, none of these metals are fitted during \cecilia's optimisation procedure.

\begin{figure*}
  \centering
  \includegraphics[width=1\linewidth]{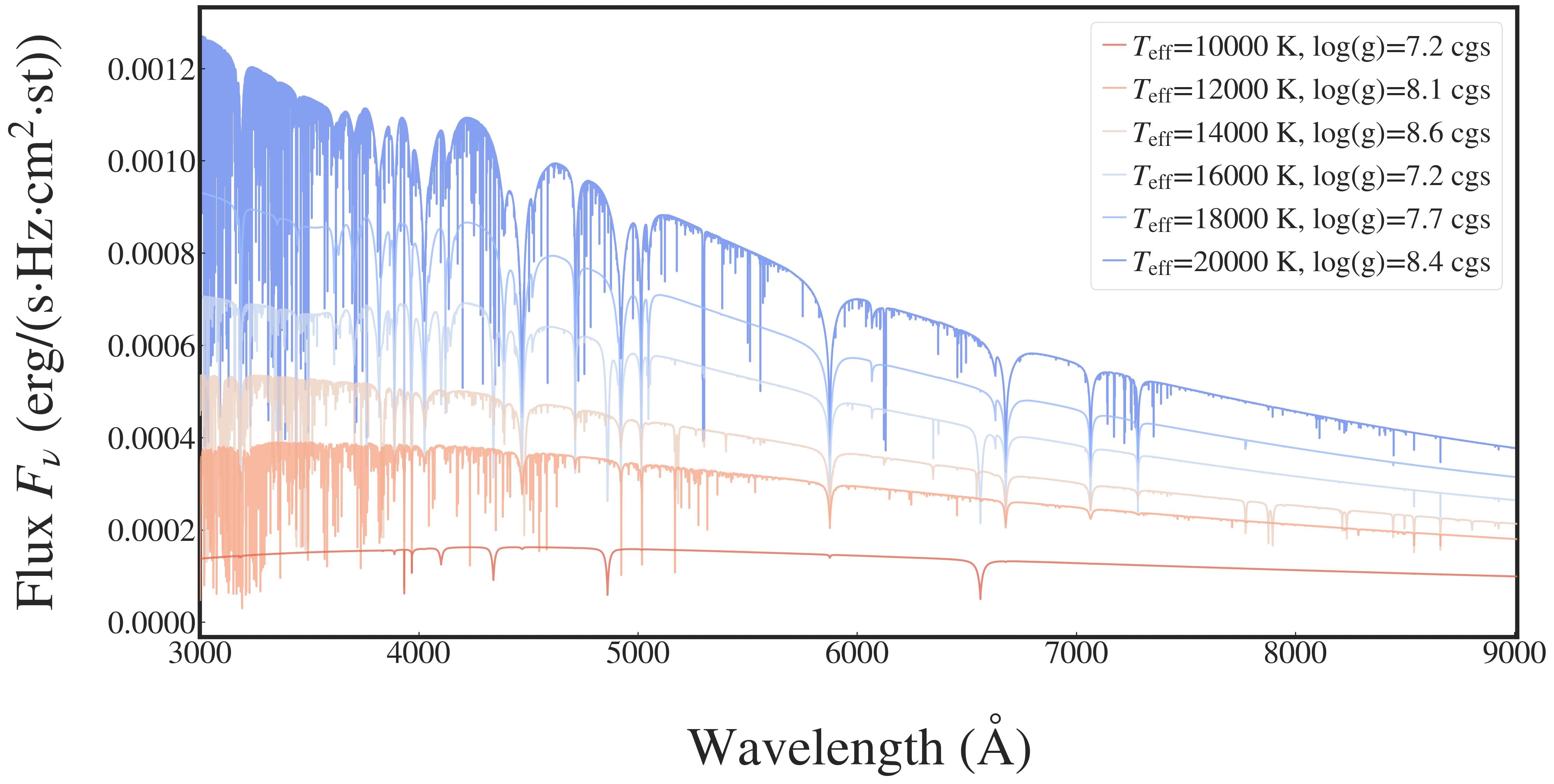}
  \caption{A selection of 6 atmosphere models for synthetic He-rich polluted white dwarfs with effective temperatures between 10,000~K and 20,000~K, and surface gravities between 7~cgs and 9~cgs. Cool and hot white dwarfs are shown, respectively, in red and blue.}
  \label{fig:synth_spectra_combined}
\end{figure*}

To prepare \cecilia's training set of synthetic spectra, we generated 25,000 unique combinations of the 13 aforementioned stellar labels (i.e. \Teff, \logg, and 11 elemental abundances). In particular, for \Teff{}, we generated our labels from a random uniform distribution spanning the range [10,000, 20,000]~K. At higher effective temperatures, the presence of photospheric metal pollution may not be due to the accretion of exoplanetary material, but to the effects of radiative levitation pressure \citep{Chayer:1995}. For cooler white dwarfs, there is less thermal energy to produce atomic transitions in the optical; as a result, there is a limited number of polluted systems exhibiting multiple heavy elements at once. With respect to \logg{}, we followed \citet{Chandra:2020} and sampled our values from a random uniform distribution between [7,9] cgs (with $g$ in cgs units of cm$\cdot$s$^{-2}$). For \logHHe, we also drew our labels from a random uniform distribution between [-7,-3]~dex, as in \citet{Coutu:2019}.

To generate our labels for the other elements, we used the following procedure. First, we drew our calcium abundances from a random uniform distribution between [-12,-7]~dex:

\begin{equation}
    \mathcal{U}_{\log(\rm Ca/He)} = \mathcal{U}(-12, -7).
    \label{eq:label_uniform}
\end{equation}
For the remaining 10 heavy elements (i.e. Mg, Fe, O, Si, Ti, Be, Cr, Mn, and Ni), we generated a random Gaussian distribution centered around $\mu_{Z}$ with a standard deviation $\sigma_{Z}$,

\begin{equation}
    \mathcal{N}_{\rm \log(Z/Ca)} = \mathcal{N}(\mu_{\rm Z}, \sigma_{\rm Z}),
    \label{eq:label_gaussian}
\end{equation}
where $\sigma_{Z}=2$~dex and
\begin{equation}
    \mu_{{\rm Z}} = \log_{10}\left({\rm Z/He}\right)_{\odot}-\log_{10}\left({\rm Ca/He}\right)_{\odot}=  \log_{10}\left({\rm Z/Ca}\right)_{\odot}.
    \label{eq:label_gaussian_mu}
\end{equation}
In the above expressions, $Z$ is the metal under consideration, $\log_{10}(\rm Z/He)_{\odot}$ is the abundance of metal $Z$ in the solar photosphere \citep{Asplund:2009}, and $\log_{10}(\rm Ca/He)_{\odot}$ is the solar calcium abundance relative to helium \citep{Asplund:2009}. We note that our choice of $\sigma_{\rm Z}=2$~dex was motivated by two factors: on the one hand, $\sigma$ had to be large enough to allow \cecilia{} to measure the true elemental abundances of a white dwarf, even if the latter deviated from chondritic values; and on the other hand, we also wanted $\sigma$ to be small enough to down-weight  unlikely regions of the parameter space in our training set, hence limiting the hypervolume of \cecilia's label parameter space. 

As a final step, we generated the distribution $\mathcal{D}$ of abundance ratios $\log(\rm {Z/He})$ for our training set by adding

\begin{equation}
    \mathcal{D} = \mathcal{N}_{\rm \log(Z/Ca)} + \mathcal{U}_{\log(\rm Ca/He)} 
    \label{eq:label_ZHe_WD}
\end{equation}
In Table \ref{tab:ranges_labels}, we list the values of $\mu_{\rm Z}$ and $\sigma_{\rm Z}$ for Mg, Fe, O, Si, Ti, Be, Cr, Mn, and Ni, rather than the properties of the (non-Gaussian) distributions resulting from evaluating Eq. \ref{eq:label_ZHe_WD}.

\begin{table}
  \centering
  \caption{\cecilia's ranges used to generate the training set of synthetic white dwarf properties (or stellar ``labels''). All the atmospheric abundances are expressed as number abundance ratios in base 10 relative to helium. References (Ref.): [1] \citet{Chandra:2020}, [2] \citet{Coutu:2019}, [3] \citet{Asplund:2009}.}
  \label{tab:ranges_labels}
  \renewcommand{\arraystretch}{1.35}
  \begin{tabular}{|lccc|}
    \hline 
    \multicolumn{4}{|c|}{\it Sampled from a Random Uniform Distribution} \\
    \hdashline[0.2pt/1.5pt]
    Label & Min. Bound & Max. Bound & Ref. \\
    \hline 
    \Teff{} [K]   & 10,000 & 20,000  &  -     \\     
    \logg{} [cgs] &   7    &  9      & [1]    \\     
    \logHHe       &  -7    & -3      & [2]    \\     
    \logCaHe      &  -12   & -7      & [2]    \\     
    \hline
    \multicolumn{4}{|c|}{\it\shortstack{Sampled from a Random Gaussian Distribution (Eq. \ref{eq:label_gaussian_mu})}}\\
    \hdashline[0.2pt/1.5pt]
    Label &  $\mu_{\rm Z}$ [dex] & $\sigma_{\rm Z}$ [dex] & Ref. \\
    \hline 
    $\log\left({\rm Mg/Ca}\right)$    &  1.26     & 2 & [3] \\      
    $\log\left({\rm Fe/Ca}\right)$    &  1.16     & 2 & [3]\\       
    $\log\left({\rm O/Ca }\right)$    &  2.35     & 2 & [3]\\       
    $\log\left({\rm Si/Ca}\right)$    &  1.17     & 2 & [3]\\       
    $\log\left({\rm Ti/Ca}\right)$    &  -1.39    & 2 & [3]\\       
    $\log\left({\rm Be/Ca}\right)$    &  -4.96    & 2 & [3]\\       
    $\log\left({\rm Cr/Ca}\right)$    &  -0.70    & 2 & [3]\\       
    $\log\left({\rm Mn/Ca}\right)$    &  -0.91    & 2 & [3]\\       
    $\log\left({\rm Ni/Ca}\right)$    &  -0.12    & 2 & [3]\\       
    \hline
  \end{tabular}
\end{table}

Finally, for the remaining 14 heavy elements (C, N, Li, Na, Al, P, S, Cl, Ar, K, Sc, V, Co, and Cu), we chose to scale their abundances such that their abundance ratios relative to calcium corresponded to the values of CI chondrites reported in in \citet{Lodders:2003}. This type of bodies are primitive carbonaceous meteorites with a pristine composition unaltered by melting and differentiation processes. With the exception of their volatile components, including their lithium and noble gas abundances, their chemical makeup ---mostly dominated by O, Mg, Si, and Fe-- closely resembles that of the solar photosphere \citep{Palme:2014} as well as that of bulk Earth \citep[e.g.][]{Zuckerman:2007, Xu:2019, Doyle:2023}. Therefore, CI chondrites are often used as a proxy for the average bulk composition of non-volatile objects in the Solar System \citep{Lodders:2003, JuraYoung:2014}. In the field of polluted white dwarfs, the assumption of chondritic ratios often constitutes a reasonable first-order approximation when there is no \textit{a priori} knowledge of the stellar photospheric composition  \citep{Dufour:2007, Coutu:2019}. In the context of this work, the use of chondritic abundances for the 14 ``fixed'' metals also allowed us to dramatically decrease the volume of the parameter space that \cecilia{} had to learn during training.

\subsubsection{Synthetic Atmosphere Models} \label{sec:synth_models}

Next, we used the 25,000 synthetic labels to generate their corresponding atmosphere models. To achieve this, we employed the code described in \citet{Dufour:2007, Blouin:2018a, Blouin:2018b} and references therein. This code includes all the necessary physical principles to predict the emerging spectra of metal-polluted white dwarfs and it has been used successfully to reproduce ultraviolet (UV) and optical observations across a wide range of effective temperatures and metal abundances \citep{Dufour:2012, Giammichele:2012, Vanderburg:2015, Melis:2017, Xu:2017, Blouin:2019a, Blouin:2019b, Coutu:2019, Blouin:2020, Kaiser:2021, Klein:2021, Caron:2023, Doyle:2023}. More specifically, for each model, the code calculates a 1D thermodynamic structure first, and this stratification is used to generate a synthetic spectrum with no instrumental broadening and sampled with a pixel spacing equal to $R\approx50,000$ (i.e. 55,000 wavelength points between 3,000~\angs{} and 9,000~\angs{}). \autoref{fig:synth_spectra_combined} shows a selection of synthetic spectra for He-rich polluted white dwarfs spanning different effective temperatures, surface gravities, and photospheric compositions.  

For \cecilia, we independently varied 13 parameters for each model (i.e. \Teff, \logg, and 11 elemental abundances; see Table \ref{tab:ranges_labels}), which we randomly selected for each calculation as described in Section \ref{sec:synth_labels}. In total, we calculated 25,000 atmosphere models, but 2,176 of them (8.7$\%$) had to be discarded due to computational issues, leaving 22,824 useful synthetic spectra for training, testing, and validation.\footnote{The missing models were not clustered in any specific region of the parameter space. Therefore, they did not negatively affect the robustness of \cecilia{}.} We note that exclusion of these models did not have any significantly impact \cecilia's predictive ability; this is due to our code's ML-based spectral interpolation approach, which facilitates the computation of spectral models in regions of the parameter space where there are no observations. We also recall that while the abundances of 10 metals (Ca, Mg, Fe, O, Si, Ti, Be, Cr, Mn, Ni) were varied independently between different synthetic spectra, the remaining 14 heavy elements (C, N, Li, Na, Al, P, S, Cl, Ar, K, Sc, V, Co, and Cu) were also included in our models with their abundance ratios Z/Ca fixed to their nominal CI chondritic abundance from \citet{Lodders:2003}.

\subsection{Data Normalisation}\label{sec:data_normalisation}

As is standard procedure in ML, we then normalised the synthetic spectra and their corresponding stellar labels between 0 and 1 in order to improve \cecilia's learning abilities during training. For our synthetic spectra, we experimented with different scaling techniques and ultimately adopted a ``pixel-wise'' (PW) normalisation approach (also commonly known as \textit{min-max} scaling in ML). For a given wavelength index $\lambda_{i}$, this approach performs the following linear transformation: first, it determines the minimum and maximum values (or ``pixels'') of the synthetic flux across \textit{all} the available spectra; and then, it uses these flux bounds to normalise the pixel at wavelength $\lambda_{i}$ to a value between 0 and 1. Therefore, the normalisation is conducted with respect to the flux axis, rather than along the wavelength dimension.

\begin{figure*}
  \centering
  \includegraphics[width=1\linewidth]{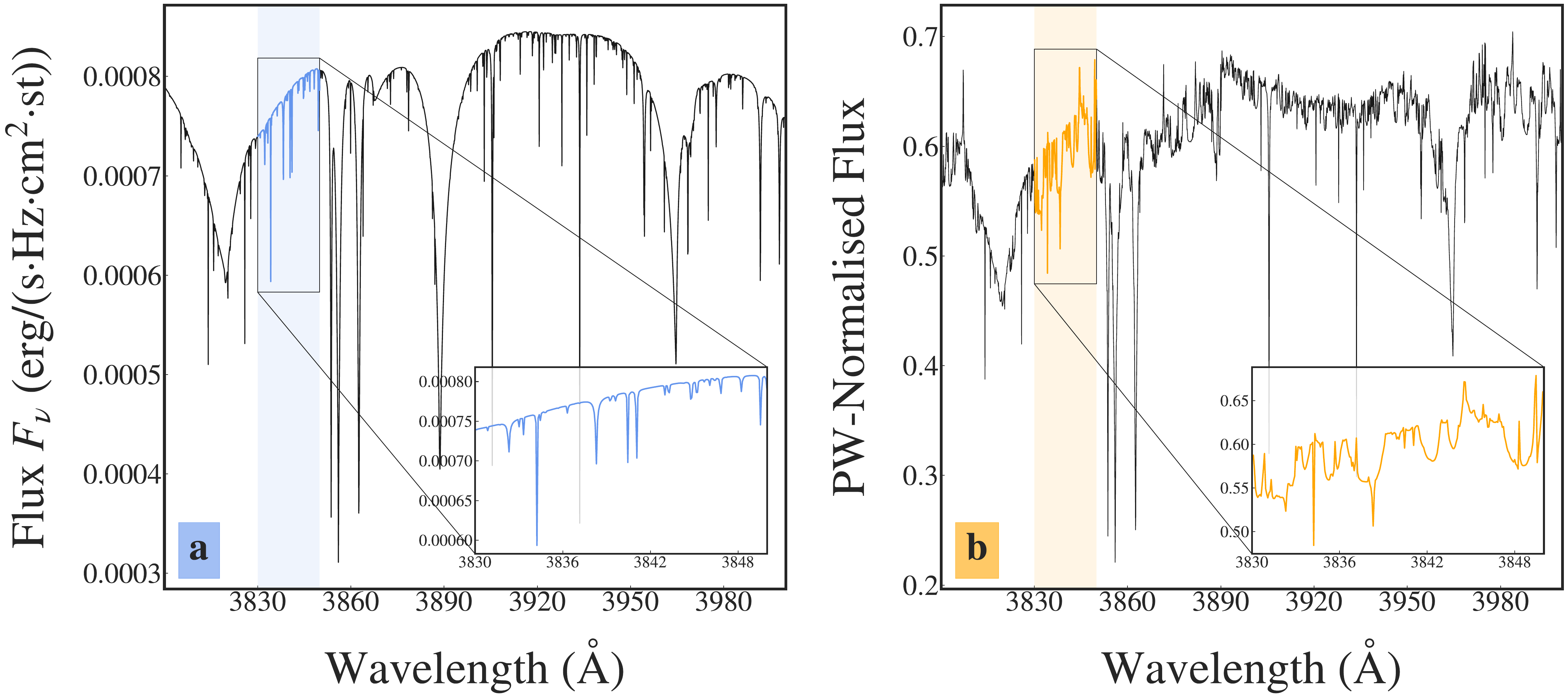}
  \caption{Effect of applying the Pixel-Wise (PW) normalisation technique on a synthetic spectrum in the spectral window between 3,800~\angs{} to 4,000~\angs. This technique is designed to improve \cecilia's learning capabilities by amplifying any small variations in the stellar flux and attenuating the largest fluctuations. Panels (a) and (b) show, respectively, the raw and PW-normalised synthetic flux of the spectrum. The inset plots show the region between 3,330~\angs{} and 3,350~\angs. }
  \label{fig:pw_norm}
\end{figure*}

As illustrated in \autoref{fig:pw_norm}, the PW approach may sometimes cause the normalised spectra to appear distorted to the human eye. Nevertheless, it is a powerful scaling technique because it treats each pixel as an independent spectral ``feature,'' thus allowing \cecilia{} to learn both the small- and large-scale variations in the synthetic flux. In other words, pixels associated to large variations become less prominent, while those linked to smaller changes become magnified. In addition to this advantage, the PW normalisation differs from other ML scaling techniques (e.g. log-scaling) because it treats each pixel in the same way, therefore ensuring that \cecilia{} does not have any preference for a specific absorption line.

\section{Methodology}\label{sec:methods}

\subsection{Design of \cecilia's ML Architecture}\label{sec:MLarchitecture}

The central problem in this paper lies in accurately determining the main properties ---or stellar labels--- of polluted white dwarfs from their spectra. To achieve this goal, we have developed \cecilia, an innovative ML pipeline capable of performing two tasks: (i) the generation of polluted white dwarf spectra from a set of stellar labels, i.e. $\mathcal{S}=\beta_{1}(\mathcal{L})$, where $\mathcal{L}$ is the vector space of the white dwarf labels (with dimension $L$), and $\beta_{1}$ is a generic non-linear function; and (ii) the prediction of polluted white dwarf labels from their corresponding spectra, or $\mathcal{L}=\beta_{2}(\mathcal{S})$, where $\mathcal{S}$ is the vector space of the white dwarf spectrum (with dimension $S$), and $\beta_{2}$ is a function different to $\beta_{1}$.

Developed with the open-source Python \texttt{tensorflow} package \citep{tensorflow2015-whitepaper}, \cecilia{} is characterised by three deep neural networks: an Autoencoder, a Fully Connected Neural Network (FCNN1), and a replica of the FCNN1 for fine-tuning purposes (FT FCNN2). The Autoencoder is a ML tool consisting of two symmetrical networks (\autoref{fig:autoencoder}): an \textit{Encoder} and a \textit{Decoder}. The \textit{Encoder} identifies the main attributes of the input data and compresses the latter into a lower-dimensional representation. This operation takes place in the so-called ``hidden'' or ``latent'' space (denoted by $\mathcal{H}$), where $\mathcal{H}$ has a dimensionality equal to the number $H$ of hidden/latent features used to encode the data (i.e. $\mathcal{H}\in\mathbb{R}^{H}$). The second part of the Autoencoder is the \textit{Decoder}, which is a network designed to reconstruct the original input data based on the \textit{Encoder}'s latent model while trying to minimising the loss of information caused by the dimensionality reduction. We emphasise that the hidden features of the Autoencoder are fundamental properties of the ML model and are not intrinsically correlated to any stellar parameters. As a result, they have no astrophysical interpretation. In general, the choice of $H$ constitutes a trade-off between the interpretability and the complexity of the ML model, characterised respectively by a low and a high $H$ \citep{Liang:2023}. 

Following the Autoencoder, \cecilia{} incorporates two deep neural networks: the FCNN1 and the FT FCNN2. As shown in Figures \ref{fig:fcnn1} and \ref{fig:fcnn2}, these networks consist of an input layer, six hidden layers, and an output layer. Unlike the Autoencoder, which is used to perform compression and dimensionality reduction, the FCNN1 and FT FCNN2 architectures are designed to teach \cecilia{} how to interpolate white dwarf spectra based on a set of 13 stellar labels. We provide a more detailed discussion about this subject in Section \ref{sec:training}. 

\begin{table}
  \centering
  \caption{Main properties of \cecilia{}'s ML architecture. Its three neural networks are trained sequentially, but only the final one (i.e. the FT FCNN2) is used to model the spectra of He-rich polluted white dwarfs.}\label{tab:ml_properties}
  \renewcommand{\arraystretch}{1.35}
  \begin{tabular}{|lc|}
    \hline
    Parameter & Value \\
    \hline \hline 
    \multicolumn{2}{|c|}{\it Autoencoder} \\
    \hline 
    No. of layers in Encoder &     5           \\
    No. of layers in Decoder &     5           \\
    No. of neurons/layer     &    4,000        \\
    No. of training epochs   &    20,000       \\
    No. of latent features   &     100         \\
    Learning rate            &  $10^{-6}$      \\
    Batch size training      &     64          \\
    Batch size validation    &     64          \\
    Activation functions     &  \makecell{ReLu \footnotesize{(initial 4 layers)} \\ sigmoid \footnotesize{(output layer)}} \\
    Weight Initialisers      &  \makecell{\texttt{he\_uniform} \footnotesize{(initial 4 layers)} \\ \texttt{glorot\_uniform} \footnotesize{(output layer)}} \\
    \hline 
    \multicolumn{2}{|c|}{\it Fully-Connected Neural Network (FCNN1)} \\
    \hline 
    No. of total layers      &     6          \\
    No. of neurons/layer     &     2,000      \\
    No. of training epochs   &    10,000      \\
    Learning rate            &  $10^{-5}$     \\
    Batch size training      &     64         \\
    Batch size validation    &     64         \\
    
    Activation functions     &  \makecell{ReLu \footnotesize{(initial 5 layers)} \\ sigmoid \footnotesize{(output layer)}} \\
    Weight Initialisers      &  \makecell{\texttt{he\_uniform} \footnotesize{(initial 5 layers)} \\ \texttt{glorot\_uniform} \footnotesize{(output layer)}} \\
    \hline 
    \multicolumn{2}{|c|}{\it Fine-Tune Fully-Connected Neural Network (FT FCNN2) } \\
    \hline 
    No. of total layers      &     6          \\
    No. of neurons/layer     &    2,000      \\
    No. of training epochs   &    1,000       \\
    Learning rate            &  $10^{-4}$     \\
    Batch size training      &     64         \\
    Batch size validation    &     64         \\
    Activation functions     &  \makecell{ReLu \footnotesize{(initial 5 layers)} \\ sigmoid \footnotesize{(output layer)}} \\
    Weight Initialisers      &  \makecell{\texttt{he\_uniform} \footnotesize{(initial 5 layers)} \\ \texttt{glorot\_uniform} \footnotesize{(output layer)}} \\
    \hline
  \end{tabular}
\end{table}

At the level of individual neurons, the choice of weights, biases, and activation functions can also shape the performance of a NN, as explained in Section \ref{sec:fundamentals_ml}. In this work, we followed common ML practice and set the weights of \cecilia{} to random values before optimising them during training. More specifically, we used the \texttt{tensorflow} functions \texttt{he\_uniform} and \texttt{glorot\_uniform} to initialise the weights of all the initial layers and the final output layer, respectively. These two functions draw values from random uniform distributions in the range between [-$\mathrm{limit}$, +$\mathrm{limit}$]. If $\mathrm{fan_{in}}$ and $\mathrm{fan_{out}}$ are, respectively, the number of input and output connections of a neuron, the parameter ``$\mathrm{limit}$'' in \texttt{tensorflow} is defined as 
\begin{equation}
    \text{limit}=\begin{cases}
    \sqrt{6/\mathrm{fan_{in}}} & (\texttt{{he\_uniform}})\\
    \sqrt{6/(\mathrm{fan_{in}}+\mathrm{fan_{out}})} & (\ensuremath{\texttt{{glorot\_uniform}}})
    \end{cases}.
\end{equation}

With respect to \cecilia's activation functions, we opted for using a Rectified Linear Unit (ReLU; \citealt{Jarrett:2009}) and a sigmoid function \citep{Goodfellow:2016}, as specified in Table \ref{tab:ml_properties}. These two activation functions regulate the flow of information by introducing non-linearities across the networks, normalising the neurons' output parameters, and enabling the identification of hidden and complex patterns in the training sample. For an input value $x$, the ReLU function returns 0 if $x$ is negative, or $x$ if $x$ is positive,

\begin{equation}
    \text{ReLU}(x) =\begin{cases}
    0 & \text{if }x<0\\
    x & \text{otherwise}
    \end{cases}={\rm max}(0,x). \label{eq:relu}
\end{equation}

Therefore, its derivative is given by 

\begin{equation}
    \frac{d}{dx}\left[\text{ReLU}(x)\right]=\begin{cases}
0 & \text{if }x\leq0\\
1 & \text{if }x>0
\end{cases}.\label{eq:relu_gradient}
\end{equation}

In contrast, the sigmoid activation function $\sigma(x)$  transforms the input value $x$ into a value between 0 and 1 with

\begin{equation}
    \sigma(x) = \frac{1}{1+e^{-x}}.\label{eq:sigmoid}
\end{equation}

As a consequence, its derivative approaches zero when $x\rightarrow \pm \infty$. In neural networks with multiple hidden layers, the potential saturation of the sigmoid function towards zero values can lead to the so-called ``vanishing gradient problem'' \citep{Hanin:2018}. This issue arises during backpropagation when the gradient of the loss function ---which encapsulates the product of the derivatives of the activation functions with respect to their weights (see Section \ref{sec:fundamentals_ml})--- becomes extremely small. Due to this ``vanishing'' effect, the weights and biases of the initial layers may only be marginally updated, which can cause poor and slow model convergence. As evidenced by Eq. \ref{eq:relu_gradient}, ReLU functions are more robust to this problem because they have a more stable derivative (either 0 or 1). As a result, they often yield a better performance during training \citep{Nair:2010_relu}. In this work, we decided to employ ReLU activation functions for the initial layers of \cecilia's networks in order to avoid the vanishing gradient of the loss function. For the final layer of each network, we chose to use a sigmoid function with the goal of generating predicted values between 0 and 1.

In Table \ref{tab:ml_properties}, we summarise the fundamental properties of \cecilia's Autoencoder, FCNN1, and FT FCNN2. These properties were determined with the Python \texttt{tensorboard} library, which is a built-in optimisation module of \texttt{tensorflow} designed to execute and compare several configurations of the same ML architecture. In particular, \texttt{tensorboard} performs a short training of each configuration to identify the one with the most optimal performance. During \cecilia's development phase, we used \texttt{tensorboard} to optimise the number of layers, the neurons per layer, the batch size (i.e. the number of training instances processed together during a given iteration), and the learning rate of the Autoencoder and the two FCNNs.

\subsection{Network Training} \label{sec:training}

\begin{figure*}
    \centering
    \begin{subfigure}
        \centering
        \includegraphics[width=0.99\linewidth]{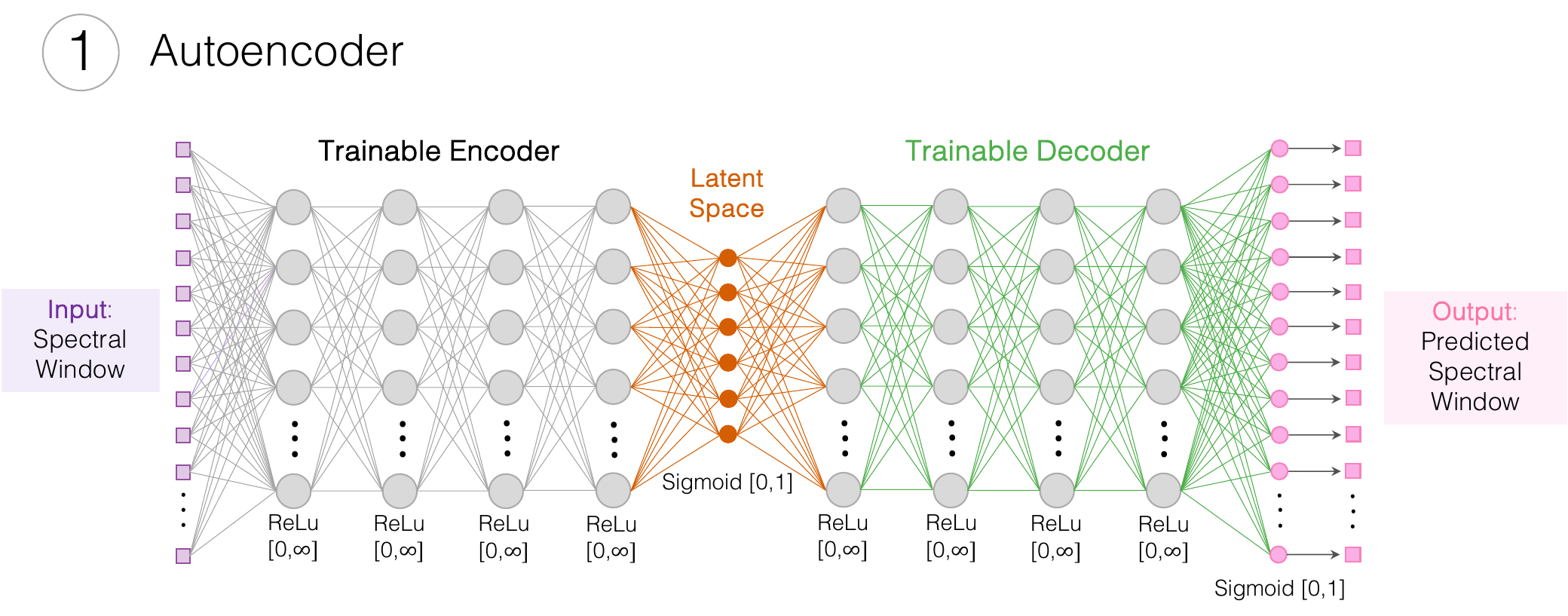}
        \caption{The first component of \cecilia's ML architecture is an Autoencoder, which consists of two networks: an Encoder (in grey)  to compress input data into a small number of latent, hidden features (in orange); and a Decoder (in green) to reconstruct the encoded data from the latent space. In our work, the input and output data of the Autoencoder are spectral windows of 200~\angs{} from our atmosphere models (in purple and pink, respectively). }
        \label{fig:autoencoder}
    \end{subfigure}
    \begin{subfigure}
        \centering
        \includegraphics[width=0.98\linewidth]{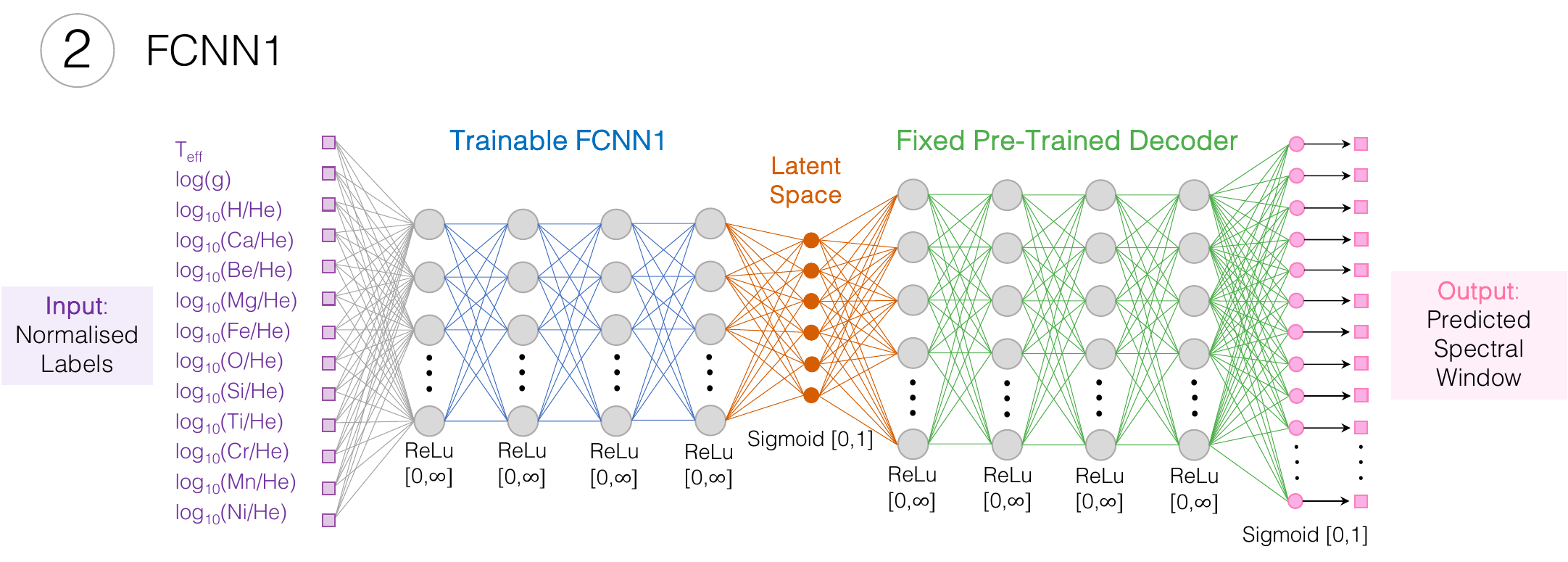}
        \caption{The second component of \cecilia's ML architecture is a feed-forward Fully Connected Neural Network (FCNN1; in blue). This network leverages the trained Decoder in \autoref{fig:autoencoder} (in green) to predict the spectrum of a polluted white dwarf (in pink) from its corresponding stellar labels (in purple). }
        \label{fig:fcnn1}
    \end{subfigure}
    \begin{subfigure}
        \centering
        \includegraphics[width=0.98\linewidth]{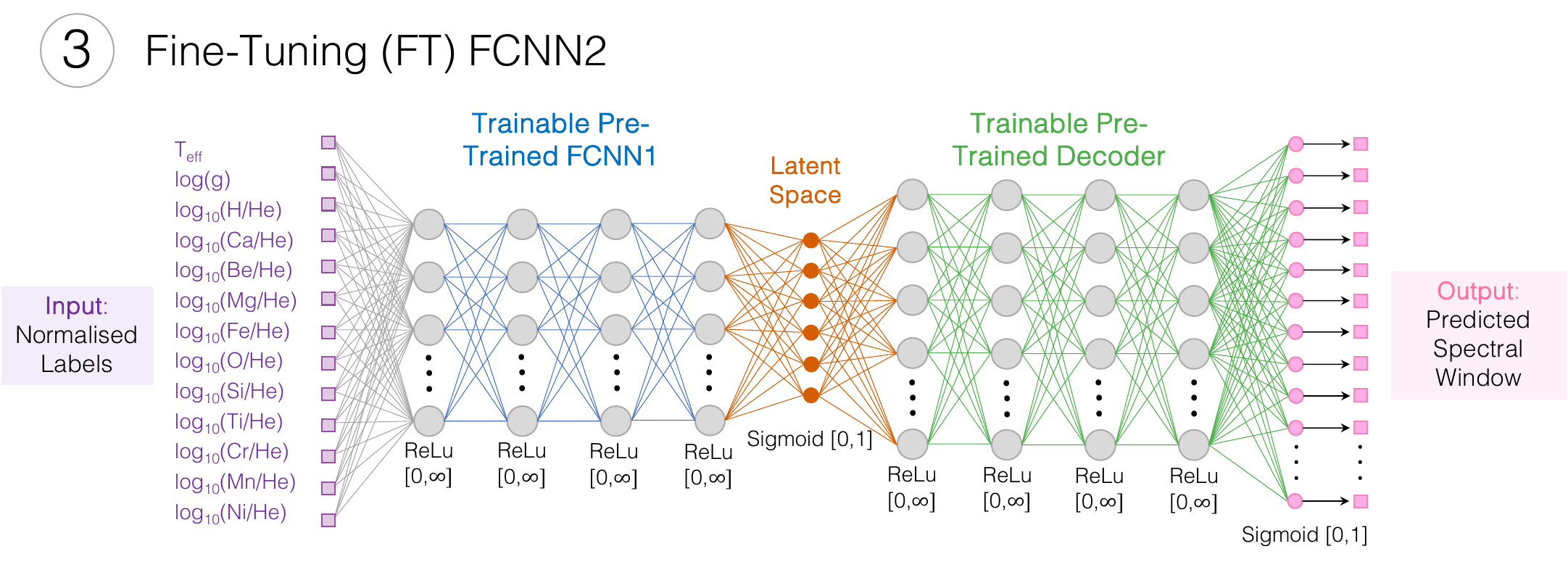}
        \caption{The third component of \cecilia's ML architecture is a replica of the FCNN1 aimed at fine-tuning the final predictions of our pipeline (FT FCNN2; in blue). Similarly to the FCNN1  
        \autoref{fig:fcnn1}, this network is connected to the trained Decoder (in green) to predict a metal-polluted spectrum (in pink) from its stellar labels (in purple).}
        \label{fig:fcnn2}
    \end{subfigure}
\end{figure*}

\begin{figure*}
      \centering
      \includegraphics[width=1\linewidth]{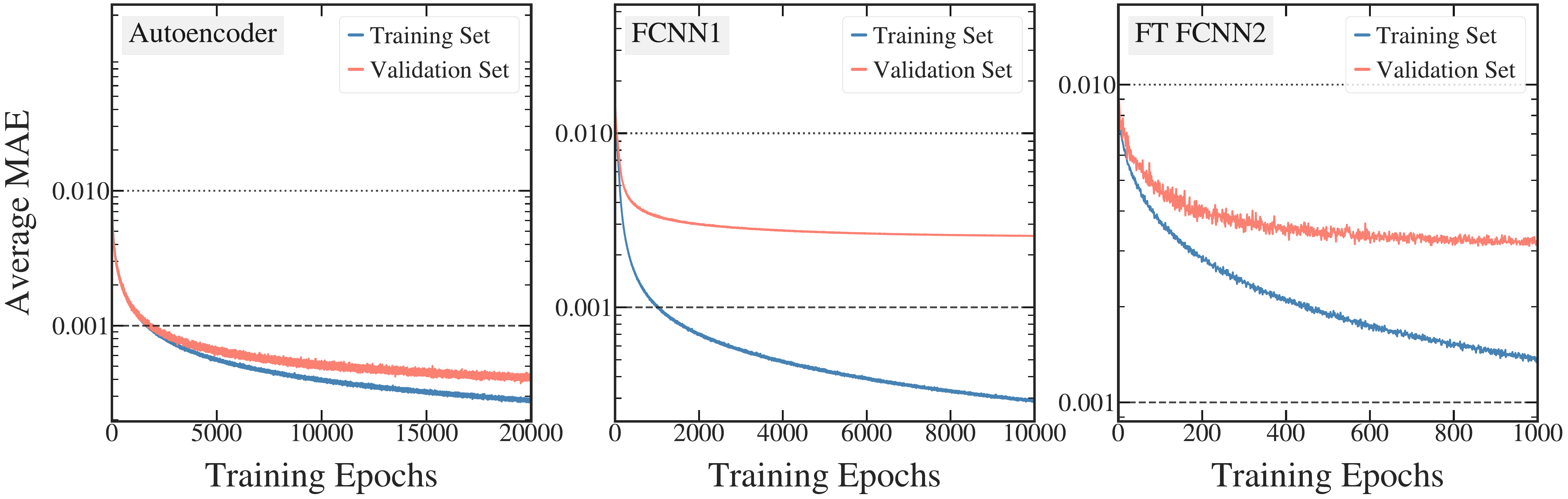}
      \caption{Mean Absolute Error (MAE) statistic (or loss function) for the Autoencoder, FCNN1, and FT FCNN2, averaged for the 29 windows of 200~\angs{} used to train \cecilia. The training and validation sets are shown in blue and red. The dotted and dashed lines denote an average MAE of 0.01 and 0.001, respectively.}
      \label{fig:mae_networks}
\end{figure*}

To train \cecilia, we used a partition of MIT's Satori Supercomputer with one NVidia V100 GPU.\footnote{\url{https://mit-satori.github.io/}.} First, we randomly split our collection of synthetic labels and spectra into three datasets for training (70$\%$), testing (20$\%$), and validation (10$\%$) purposes. Next, we used the backpropagation technique to optimise the hyperparameters of \cecilia's neural networks. To this end, we implemented the Adaptive Moment Estimation (Adam) optimisation algorithm \citep{Adam:2017}, which is an extension of the commonly used Stochastic Gradient Descent routine \citep{Kiefer:1952}. Finally, given that the synthetic spectra described in Section \ref{sec:synth_models} represented a substantial amount of data (a total of about $10^{9}$ flux pixels for 22,842 spectra), we chose to train our pipeline by dividing the synthetic spectra in 29 windows of 200~\angs, rather than by considering the full wavelength range at once.\footnote{We also evaluated the performance of \cecilia{} with narrower and wider spectral windows. From this analysis, we found that 200~\angs{} offered the highest computational efficiency and predictive accuracy.} With this approach, the end-to-end training of \cecilia{} took approximately three weeks. 

For each spectral window of 200~\angs, \cecilia's training consisted of three sequential learning phases: an initial one for the Autoencoder, another one for the FCNN1, and a final one for the FT FCNN2. In each phase, the networks employed the same datasets for training, validation, and testing purposes, with each dataset consisting of unique combinations of the 13 randomly generated labels listed in Table \ref{tab:ranges_labels} (i.e. \Teff, \logg, and 11 elemental abundances from H to Ni). It is important to clarify that the Autoencoder and the FCNN1 were \textit{only} used to improve the learning of the fine-tuned neural network. After concluding \cecilia's training, our code only uses the FT FCNN2 to generate spectral models of polluted white dwarfs. Below, we summarise the three training phases in more detail.

\begin{itemize}[noitemsep,topsep=1pt,leftmargin=12pt]
    \item \textit{Phase 1 (\autoref{fig:autoencoder})}: First, we trained the Autoencoder in order to compress and reconstruct the normalised synthetic spectra based on only 100 latent features. At the end of this process, we used the trained Encoder to transform our full database of 22,824 synthetic spectra into their latent space representations. If $\mathcal{S}$ and $\mathcal{H}$ represent, respectively, the vector spaces of the synthetic spectra and the hidden space (see Section \ref{sec:MLarchitecture}), we can denote the operations of the trained Autoencoder as
    \begin{equation}
        \mathcal{S} {\rightarrow} \mathcal{H} {\rightarrow} \mathcal{S}.
        \label{eq:autoencoder_operation}
    \end{equation}
    \item \textit{Phase 2 (\autoref{fig:fcnn1})}: Next, we trained the first Fully-Connected Neural Network (FCNN1) to compress the normalised stellar labels into their latent space representations. We then appended the trained Decoder to the output layer of the FCNN1 to convert the latent features of the normalised labels into predictions of normalised synthetic spectra.  If $\mathcal{L}$ is the vector space of the stellar labels, the FCNN1 and trained Decoder are thus in charge of perfoming the operation:
    \begin{equation}
        \mathcal{L} {\rightarrow} \mathcal{H} {\rightarrow} \mathcal{S}.
        \label{eq:fcnn1_operation}
    \end{equation}
    
    At this stage, we had two options for training the FCNN1 while keeping the trained Decoder fixed. On the one hand, we could directly determine the errors of the hidden layer with respect to the encoded spectra $\mathcal{H}$. This scenario required prior compression of the spectra with the Encoder and involved backpropagating the error through the FCNN1 to correct the weights and biases of the network. On the other hand, we could calculate the errors of the FCNN1's output layer with respect to the true synthetic spectra $\mathcal{S}$. This alternative option required backpropagating the errors through the fixed Decoder until reaching the hidden space, and then updating the weights and biases of the FCNN1 accordingly. In other words, the main difference between the two approaches was the location of the network used to calculate the loss function, with the first and second scenarios employing, respectively, the encoded synthetic spectra $\mathcal{H}$ and the normalised synthetic spectra $\mathcal{S}$. In this work, we decided to follow the first approach in order to (i) avoid  unnecessary backpropagation steps while traversing the Decoder, and (ii) prevent gradient vanishing effects of the loss function, which could result in a slower and poorer performance.

    \item  \textit{Phase 3 (\autoref{fig:fcnn2})}: Lastly, we sought to improve and fine-tune \cecilia's final predictions by training a second Fully-Connected Neural Network (FT FCNN2) connected to the trained Decoder. This architecture replicates the structure of the FCNN1, but it   differs from the latter because we allowed the parameters of the Decoder to be trainable. Similarly to Eq. \ref{eq:fcnn1_operation}, the joint FT FCNN2 and Trainable Decoder architecture performs the operation
    \begin{equation}
        \label{eq:ft_operation}
        \mathcal{L} {\rightarrow} \mathcal{H^{*}} {\rightarrow} \mathcal{S},
    \end{equation}
    where the latent space of this third network, represented by the term $\mathcal{H}^{*}$ in Eq. \ref{eq:ft_operation}, is now different than those of the Autoencoder and the FCNN1 ($\mathcal{H}$).
\end{itemize}

\begin{figure*}
      \centering
      \includegraphics[width=1\linewidth]{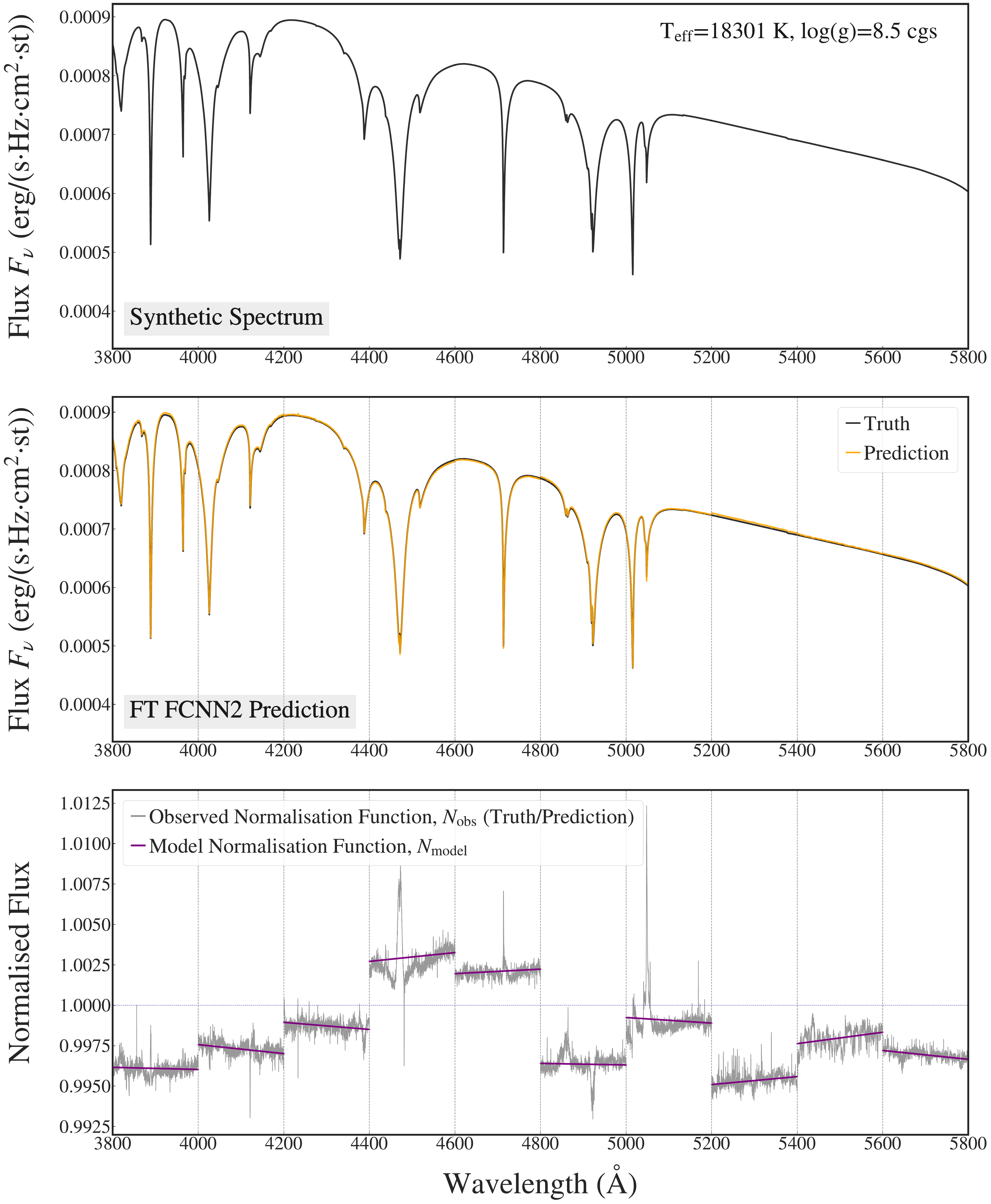}
      \caption{\textit{Top Panel}: Synthetic spectrum corresponding to a white dwarf with \Teff=18,301~K and \logg=8.5~cgs. To allow for a closer examination, we only show the wavelength range between 3,800~\angs{} and 5,800~\angs, which features 10 windows of 200~\angs. \textit{Middle Top Panel}: Denormalised \cecilia{} FT FCNN2 prediction. \textit{Bottom Top Panel}: Slope correction with a linear least-squares fit. In each panel, we use dashed vertical lines to indicate the start/end of a 200~\angs{} spectral window.}
      \label{fig:training_predictions}
\end{figure*}

\vspace{5pt}
Upon training \cecilia's Autoencoder and FCNNs, we calculated the error between the predictions of each network and their respective true synthetic models. To this end, we adopted the Mean Absolute Error (MAE) metric as our loss function ---a popular choice in ML regression problems. In this work, we calculate the MAE of a spectral window as
\begin{equation}
    {\rm MAE}=\frac{1}{Q P}\sum_{q=1}^{Q} \sum_{p=1}^{P} \left|y_{q,p}-\hat{y}_{q,p}\right|,
    \label{eq:mae}
\end{equation}
where $q$ and $Q$ represent, respectively, the index and the total number of instances ---where we recall that an instance is a synthetic spectrum with its corresponding stellar labels---, $p$ and $P$ denote the index and total number of output pixels (i.e. the number of flux points in a given spectral window), $y$ and $\hat{y}$ correspond to the true and predicted synthetic flux. \autoref{fig:mae_networks} shows our loss function for each network, with the MAE averaged for the 29 windows in the wavelength range between 3,000~\angs{} and 9,000~\angs. From this figure, we find that the training error rates of \cecilia's networks decrease with each epoch, as should be the case during training. This behavior demonstrates the ability of \cecilia{} to recognise abstract features in the training data and generalise its acquired knowledge to previously unseen observations. At the same time, \autoref{fig:mae_networks} shows that none of \cecilia's networks are suffering from the dangers of underfitting or overfitting:\footnote{The concept of \textit{underfitting} refers to a neural network that has not managed to learn the fundamental attributes of a training sample. This problem can arise from various reasons ---e.g. an excessively simple ML model, a small training set, a short training duration--- and it leads to poor predictions of new datasets. In contrast, \textit{overfitting} describes a network that has memorised the underlying features of the training data too well but cannot make accurate predictions with a test dataset. This behaviour can also happen due to multiple reasons, such as an overly complex ML model or an unnecessarily long training process. An optimal ML model represents a balance between generalisation and specificity \citep{Goodfellow:2016}.} on the one hand, the absence of underfitting can be demonstrated with the convergence of the validation error, which becomes asymptotically horizontal, thus leaving no substantial margin for further generalisation; on the other hand, the problem of overfitting can be discarded because the validation error does not increase over time. In \autoref{fig:training_predictions}, we also present an example of \cecilia's predictions for several 200~\angs{}-windows of a synthetic spectrum. 

A general note about our training process is that Phases 2 and 3 \textit{recycle} the trained Decoder to improve the performance of the FCNN1 and the FT FCNN2. After experimenting with less sophisticated ML architectures, we found this approach resulted in more accurate predictions than a simpler NN designed to predict stellar labels directly from their spectra without any intermediary steps (i.e. $L=f(S)$). The use of a pre-trained model to improve the training of another network is known as ``Transfer Learning'' and has been very successful in many scientific fields outside computational science \citep[e.g.][]{Tan:2018}.

\subsection{Parameter Estimation} \label{sec:parameter_estimation}

After training, \cecilia{} can use its FT FCNN2 architecture to rapidly ($<$1~second) generate a full high-resolution synthetic spectrum from 13 stellar labels (\Teff, \logg, \logHHe, and 10 metal abundances relative to helium). Exploiting the fast and automated ML interpolation capabilities of our pipeline, we have developed a new fitting framework to simultaneously and accurately measure the main physical and chemical properties of He-rich polluted white dwarfs from their spectra. This section explains our methodology for achieving this goal using a step-by-step approach. In our discussion, we will employ the subscripts ``synth'' and ``obs'' to represent \cecilia's synthetic predictions and a real white dwarf spectrum, respectively. With this choice of notation, we will refer to their corresponding wavelength, flux, and flux error arrays as $\lambda_{\rm X}$,  $f_{\rm X}$, and $f_{\rm X, err}$, where $X$ will indicate the nature of the data (i.e. synthetic ``synth,'' or observed ``obs''). Finally, we will denote their resolving power as $R_{\rm X}$, where $R_{\rm X}\equiv\lambda_{\rm X}/\Delta\lambda_{\rm X}$ (unitless). 

Broadly speaking, our fitting framework can be divided into three main phases (see \autoref{fig:fitting}). First, \cecilia's trained FT FCNN2 architecture generates a preliminary atmosphere model based on the assumed astrophysical properties of the white dwarf. Second, our pipeline optimises this initial prediction with the non-linear least-squares Levenberg-Marquardt algorithm implemented by the fast Python \texttt{mpfit}\footnote{\url{https://github.com/segasai/astrolibpy/blob/master/mpfit/mpfit.py}} library \citep{More:1978, Markwardt:2009}. To conclude, we perform a global Bayesian fit with the differential evolution Markov Chain Monte Carlo (MCMC) sampler of the Python \texttt{edmcmc} package \citep{Vanderburg:2021_edmcmc, Ter:2006}.\footnote{Differential evolution is a genetic algorithm designed to find an optimal distance and direction for a jumping distribution from one MCMC chain to another.} More specifically, our optimisation procedure consists of the following phases:

\begin{enumerate}[noitemsep,topsep=1pt,leftmargin=14pt,listparindent=1.5em]
    \vspace{5pt}
    \item \label{sec:step1} \textbf{Step 1: Input Spectrum and Stellar Parameters}: First, the user uploads the spectrum of their polluted white dwarf (\wobs, \fobs, \ferrobs) into \cecilia, ensuring that the effective temperature and surface gravity of the star satisfy the ranges 10,000$\leq$\Teff$\leq$20,000~K and 7$\leq$\logg$\leq$9~cgs imposed in Section \ref{sec:synth_labels} (see Table \ref{tab:ranges_labels}). In parallel, the user provides a list of 14 white dwarf properties to our pipeline, either complete with initial guesses, or lacking the latter if unknown. These properties consist of the 13 stellar labels used during \cecilia{}'s training,\footnote{A future improvement to \cecilia{} will be to include a trained neural network to estimate the effective temperature and surface gravity of the polluted white dwarf from existing photometric observations (see Section \ref{sec:discussion}). In the meantime, we assume that the user has good external constraints on \Teff{} and \logg{}.} as well as an additional Radial Velocity shift (or RV thereafter) to account for the barycentric motion and gravitational redshift of the white dwarf. If necessary, the RV term can also be used to handle potential wavelength calibration issues in the spectrum.

    \vspace{5pt}
    \item \label{sec:step2} \textbf{Step 2: Initial Guesses for Stellar Parameters}: Next, \cecilia{} reviews the user's white dwarf properties and performs the following operations based on the availability (or absence) of initial guesses for the 13 stellar labels and the RV term:
    \begin{itemize}
        \item \textit{Scenario 1: The user has prior knowledge on all stellar parameters}: In this optimal situation, \cecilia{} examines the user's initial guesses to confirm that they are within the ML bounds listed in Table \ref{tab:ranges_labels}. If the RV shift is not within the range given by -500$\leq$RV$\leq$500~km/s, our code automatically sets this parameter to 0~km/s. Alternatively, if any of the abundances are outside of their corresponding ML ranges, \cecilia{} attempts to correct this problem by generating a normal distribution centered on the user's estimate with a scale of 0.1~dex, and subsequently adopting a random value from this distribution as an initial guess. If this modification is still insufficient, \cecilia's Gaussian estimate is then substituted by a chondritic abundance ratio based on the assumed \logCaHe{} of the white dwarf and the chondritic values of \citet{Asplund:2009}. 

        \item \textit{Scenario 2: The user has partial knowledge of some stellar parameters}: In this intermediate scenario, our code starts by verifying that the user's initial guesses for \Teff{}, \logg{}, \logHHe{} and \logCaHe{} are within \cecilia's allowed bounds in Table \ref{tab:ranges_labels}. If any of these parameters are unknown, \cecilia{} uses the mean of their ML bounds as initial guesses for its optimisation routine (i.e. $\bar{T}_{{\rm eff}}$=15,000~K, $\bar{\log({\rm g}})$=8~cgs, $\bar{\log({\rm H/He)}}$=-5, and $\bar{\log({\rm Ca/He)}}$=-9.5). For the remaining white dwarf parameters, \cecilia{} adopts a zero RV shift and assumes chondritic abundances for Mg, Fe, O, Si, Ti, Be, Cr, Mn, Ni based on the estimated \logCaHe{} of the white dwarf and the chondritic ratios of \citet{Asplund:2009}.
        
        \item \textit{Scenario 3: The user has no knowledge of the stellar parameters}: In this worst-case scenario, \cecilia{} generates preliminary estimates of \Teff{}, \logg{}, \logHHe{} and \logCaHe{} by taking the mean of their corresponding ML bounds, as in Scenario 2 above. For the remaining white dwarf properties, it also follows the same procedure described in Scenario 2, that is: it assumes a RV shift of 0~km/s and chondritic abundance ratios for those elements lacking a user-defined initial guess. 
    \end{itemize}

    \vspace{5pt}
    \item \label{sec:step3}  \textbf{Step 3: Definition of \cecilia{}'s Spectral Model}: After loading the observed spectrum into \cecilia{} together with (adjusted) preliminary estimates of the 14 white dwarf parameters, the user is asked to choose the free and fixed parameters of their \cecilia{} spectral model. If certain parameters are fixed, our code freezes their initial value to the user's guess or, if not provided, to \cecilia's estimated value. 

    \vspace{5pt}
    \item \label{sec:step4} \textbf{Step 4: Adjusting the Format of the Input Data}: The last step before initiating our fitting procedure is to adjust the white dwarf spectrum and its stellar labels to a format that \cecilia{} can understand. To this end, our code automatically implements the following data processing techniques: 

    \begin{itemize}
        \item \textit{Normalisation of Initial Guess Labels}: First, it normalises the guess stellar labels using the minimum and maximum values of the synthetic labels considered during training (see Section \ref{sec:data_normalisation}). Given that \cecilia{} was trained in normalised space, this step is crucial to ensure that our code can effectively process and interpret the user's input data.
        
        \item \textit{Check of Spectral Units}: Next, \cecilia{} verifies that the observed stellar flux is expressed in terms of the calculated flux at the surface of the white dwarf, i.e. $F_{\nu}$, where $[F_{\nu}]$=erg/(s$\cdot$Hz$\cdot$cm$^{2}\cdot$sr). Although this check is not of paramount importance due to the slope and offfset correction discussed in step \ref{sec:step6} below, $F_{\nu}$ is the unit of choice of \cecilia's atmosphere models (e.g. see  \autoref{fig:training_predictions}) and is therefore the preferred notation of our pipeline. In reality, however, real spectra of white dwarfs are typically not expressed in units of $F_{\nu}$, but in $f_{\lambda}$, where $[f_{\lambda}]$=erg/(s$\cdot$cm$^{2}\cdot$\angs). As a consequence, our code integrates the possibility of transforming the observed stellar flux from $f_{\lambda}$ to $F_{\nu}$ with

        \begin{equation}
        F_{\nu}=\left(\frac{f_{\lambda}\cdot w_{{\rm obs}}^{2}}{c}\right)\cdot\left[\frac{1}{4}\left(\frac{d}{r}\right)^{2}\right],
        \label{eq:unit_conversion}
        \end{equation}
        where $c$ is the speed of light  (in [\angs/s]), $r$ is the radius of the white dwarf, $d$ is its distance from Earth, and the second term represents the inverse solid angle of the white dwarf (in ${\rm [1/sr]}$). In Section \ref{sec:results}, we use the \Gaia{} DR3 database \citep{Gaia:2016, Gaia:2023} to determine the values of $r$ and $d$. 
        \item \textit{Spectrum Cropping}: Lastly, our code examines the observed wavelength array to make sure that it falls within \cecilia's spectral coverage between 3,000~\angs{} and 9,000~\angs{}. Any regions outside these boundaries are automatically excluded from our fitting analysis (e.g. see \autoref{fig:sdss_rawspec_wd1232}).
    \end{itemize}

    \vspace{5pt}
    \item \label{sec:step5} \textbf{Step 5: \cecilia's High-Resolution Spectral Prediction}: Next, our code feeds the normalised stellar labels into \cecilia's trained FT FCNN2 architecture to quickly generate a normalised synthetic spectrum for the white dwarf (\wsynth, \fsynth). At this stage, \cecilia's synthetic spectrum shares the resolving power of the atmosphere models described in Section \ref{sec:synth_models}, i.e. \Rsynth$\approx$50,000. Therefore, assuming that the resolving power of the observed white dwarf spectrum is lower than that of our atmosphere models (\Robs$\ll$\Rsynth), \cecilia's prediction cannot be optimised yet.

    \vspace{5pt}
    \item \label{sec:step6} \textbf{Step 6: Processing of \cecilia's High-Resolution Spectral Prediction}:  To enable a meaningful comparison between the output of \cecilia's FT FCNN2 architecture and the observed input spectrum, our code performs the following automated and sequential operations: 
    
\begin{figure*}
  \centering
  \includegraphics[width=1\linewidth]{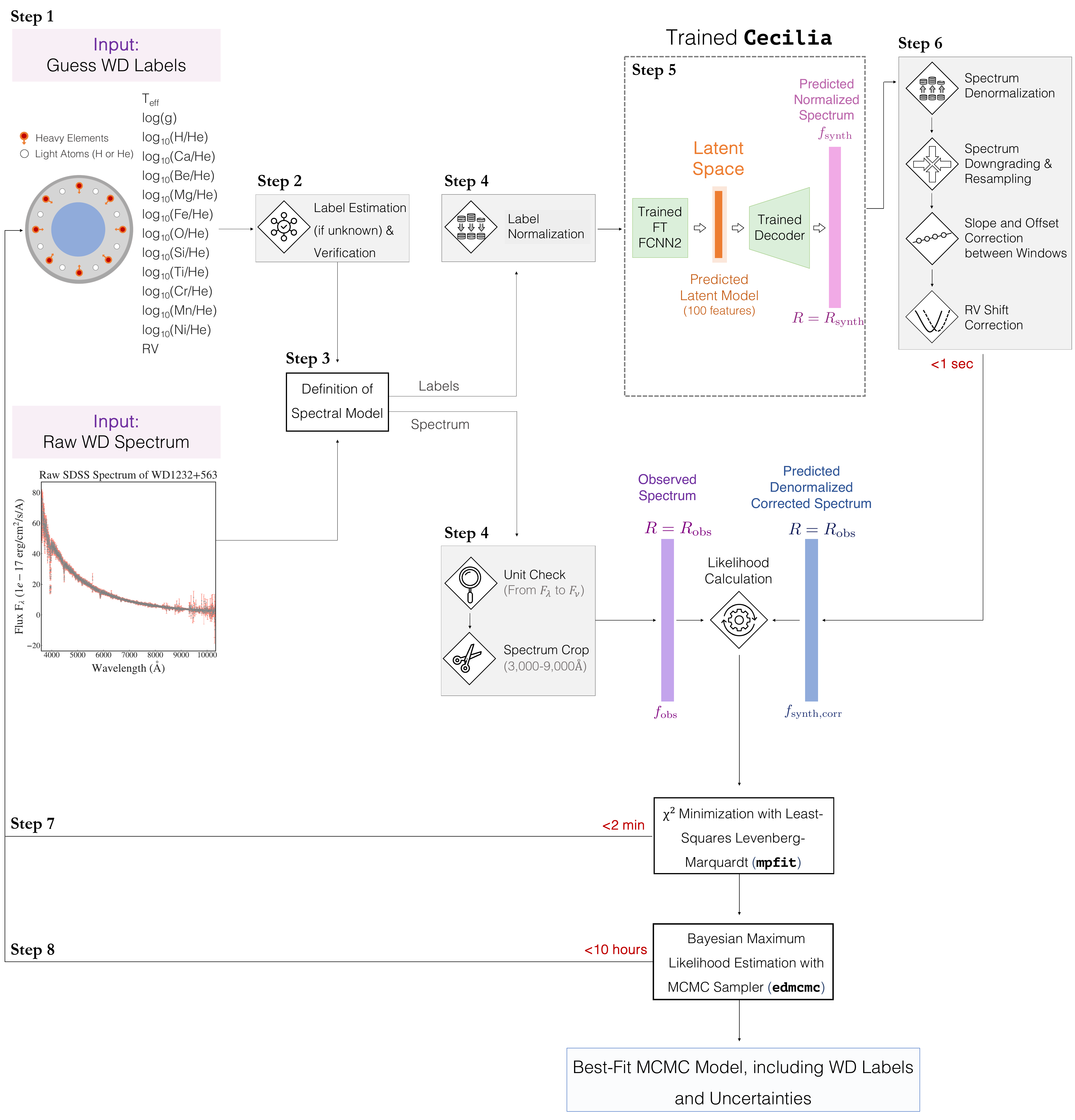}
  \caption{A schematic view of our fitting methodology using \cecilia{} and a combination of frequentist and Bayesian statistical techniques. First, the user uploads the spectrum of their polluted white dwarf onto \cecilia, together with a list of 14 stellar properties (\Teff, \logg, 11 elemental abundances, and a RV shift; see step \ref{sec:step1} in Section \ref{sec:parameter_estimation}). If these properties are unknown, \cecilia{} estimates initial guesses as explained in step \ref{sec:step2}; afterwards, it verifies that the labels are within \cecilia's ML bounds. Next, the user defines the free and fixed parameters of their spectral model (step \ref{sec:step3}). Then, our code adapts the input spectrum and stellar properties to \cecilia's format (step \ref{sec:step4}); this includes normalising the guess labels and excluding any regions of the observed spectrum outside of \cecilia's coverage between 3,000~\angs{} to 9,000~\angs{}. After this data processing step, our code feeds the normalised labels to the trained FT FCNN2 network, which then takes $<$1~second to generate a normalised, high-resolution synthetic spectrum (step \ref{sec:step5}). This prediction is subsequently denormalised and adjusted to the properties of the input spectrum (step \ref{sec:step6}) before being optimised with two techniques: a non-linear least-squares $\chi^{2}$ minimisation algorithm (step \ref{sec:step7_mpfit}), and a Bayesian MCMC (step \ref{sec:step8_mcmc}). Using 1 GPU from the MIT Satori Supercomputer, our fitting procedure produces a full spectroscopic solution in less than about 10 hours, including the stellar labels of the polluted white dwarf with their corresponding uncertainties. Source of icons: \textit{Flaticon}. }
  \label{fig:fitting}
\end{figure*}

    \begin{itemize}
        \item \textit{Denormalisation of \cecilia's Synthetic Prediction}: First, it denormalises \cecilia's prediction  with the minimum and maximum flux values of the synthetic models used during training. This transformation brings \cecilia's synthetic spectrum to the non-normalised space of the observed white dwarf spectrum. 
        \item \textit{Smearing and Resampling}: Second, our code smears and reasmples \cecilia's denormalised prediction to the resolving power (\Robs{}) and wavelength grid (\wobs{}) of the input spectrum, respectively. To achieve this, it downgrades \cecilia's synthetic spectrum from \Rsynth{} to \Robs{} by convolution with a Gaussian smoothing function. Then, it resamples \cecilia's prediction by smearing the latter onto the wavelength grid of the observed spectrum using \texttt{Python}'s one-dimensional linear interpolator \texttt{numpy.interp}. 
        
        \item \textit{Slope and Offset Correction}: Third, we correct for any spectral ``jumps''  in \cecilia's prediction caused by the training of our ML pipeline in independent windows of 200~\angs{} (e.g. see \autoref{fig:training_predictions}). For this task, we start by defining an \textit{observed} normalisation function $\mathcal{N_{\rm obs}}$ from the ratio between \cecilia's prediction and the observed flux, 
        
        \begin{equation}
        \mathcal{N_{\rm obs}}^{i} = \frac{f_{{\rm synth}}^{i}}{f_{{\rm obs}}^{i}},
        \label{eq:Nobs}
        \end{equation}
        where $f^{i}_{{\rm synth}}$ is \cecilia's synthetic flux at wavelength $i$, and $f^{i}_{{\rm obs}}$ is the observed stellar flux. Then, we use a weighted linear least-squares algorithm to fit  $\mathcal{N_{\rm obs}}$ as a function of the observed wavelength array, \wobs. From this analysis, we obtain a \textit{model} normalisation function $\mathcal{N_{\rm model}}$ (e.g. see purple line in the second panel of \autoref{fig:training_predictions}), which we then multiply to \cecilia's prediction to minimise the jumps between consecutive windows. We note that \cecilia{} can do a slope and offset correction with an \textit{n}-th degree polynomial as well.
        
        \item \textit{Radial Velocity Shift}: Finally, our code applies a radial velocity shift to \cecilia's denormalised, smeared, resampled, and slope-corrected prediction. In particular, if $c$ is the speed of light, and \rvobs{} is the assumed RV shift of the white dwarf, the total wavelength shift of the spectral lines can be calculated with

        \begin{equation}
        \delta\lambda_{\rm synth}=\frac{\lambda_{\rm synth}\cdot v_{\rm obs}}{c} \equiv \lambda_{\rm synth}^{\prime} - \lambda_{\rm synth}. 
        \end{equation}
        
        From the above, we determine the shifted wavelength array of \cecilia's synthetic prediction with
        \begin{equation}
        \lambda_{{\rm synth}}'=\lambda_{{\rm synth}}\cdot\left(1+\frac{v_{{\rm obs}}}{c}\right) \label{eq:shifted_wav}.
        \end{equation}
        
        Using this new wavelength grid, we then perform a cubic spline interpolation to obtain the shifted flux and flux error arrays (Eq. \ref{eq:shifted_flux}). In the absence of flux points within a certain wavelength range, our function simply extrapolates the data. Therefore, for a shifted wavelength array $\lambda^{\prime}_{{\rm synth}}$, the final products of our interpolation are:
        
        \begin{equation}
            \begin{array}{ccc}
            f_{{\rm synth}}(\lambda_{{\rm synth}}) & \rightarrow & f_{{\rm synth}}^{\prime}(\lambda_{{\rm synth}}^{\prime})\\
            f_{{\rm err,synth}}(\lambda_{{\rm synth}}) & \rightarrow & f^{\prime}_{{\rm err,synth}}(\lambda_{{\rm synth}}^{\prime})
            \end{array}.
            \label{eq:shifted_flux}
        \end{equation}

        In the remaining discussion, we will denote \cecilia's smeared, resampled, slope-corrected, and RV-shifted synthetic wavelength and flux as \wsynthcorr{} and \fsynthcorr{}, respectively. 
    \end{itemize}

    \item \label{sec:step7_mpfit}  \textbf{Step 7: Initial \texttt{mpfit} $\chi^{2}$ Optimisation}: After concluding steps \ref{sec:step1}-\ref{sec:step6}, our code initiates the first phase of our optimisation procedure, namely: a preliminary non-linear least-squares fit implemented with the \texttt{mpfit} package. The  goal of \texttt{mpfit} is to quickly ($<2$ minutes) identify good values for the model parameters which can then be used as initial guesses for our MCMC. To achieve this, \texttt{mpfit} minimises the chi-squared statistic ($\chi^{2}$) of \cecilia's corrected prediction, 

    \begin{equation}
        \chi^{2}=\sum_{i=1}^{N_{{\rm points}}}\left[\frac{f_{{\rm obs}}^{i}-f_{{\rm synth, corr}}^{i}}{f_{{\rm obs,err}}^{i}}\right]^{2},
        \label{eq:chi2_mpfit}
    \end{equation}
    where $N_{\rm points}$ is the total number of points in the observed input spectrum, $f^{i}_{{\rm synth, corr}}$ is \cecilia's corrected synthetic flux at wavelength $i$, $f^{i}_{{\rm obs}}$ is the observed stellar flux, and $f_{{\rm obs,err}}^{i}$ is the uncertainty of the latter. 
    
    As illustrated in \autoref{fig:fitting},  \texttt{mpfit} computes the $\chi^{2}$ metric in Eq. \ref{eq:chi2_mpfit} as follows: first, it provides the normalised stellar labels to \cecilia{}, which then invokes its trained FT FCNN2 architecture to rapidly generate a preliminary normalised, high-resolution (\Rsynth), synthetic spectrum (\wsynth, \fsynth). Next, \cecilia's prediction is adjusted with the techniques described in step \ref{sec:step6}, which allows \texttt{mpfit} to compare the observed input spectrum to \cecilia's corrected prediction. From this comparison, \texttt{mpfit} calculates the $\chi^{2}$ metric with Eq. \ref{eq:chi2_mpfit} and repeats the entire optimisation procedure until identifying a set of optimal model parameters. For these best-fit labels, \texttt{mpfit} also provides their covariance matrix, which can then be used to derive initial estimates of their uncertainties.

    \vspace{5pt}    
    \item \label{sec:step8_mcmc} \textbf{Step 8: Final \texttt{MCMC} Exploration} The second phase of our fitting routine employs the best-fit results of \texttt{mpfit} to initialise an MCMC sampler and obtain more robust parameter uncertainties. In contrast to \texttt{mpfit}, which uses frequentist (or ``classical'') statistics to minimise the $\chi^{2}$ value of \cecilia's predictions and determine best-fit values with fixed error bars, an MCMC relies on Bayesian inference to maximise the likelihood $\mathscr{L}$ of the model parameters and evaluate their full posterior probability distributions. These ``posteriors'' are proportional to the information contained in the observations as well as to any pre-conceived beliefs or assumptions about the model parameters (or ``priors''). Mathematically, the posterior probability ($\pi$) of observing the model parameters ($\boldsymbol{\theta}$) given an observed spectrum (\fobs{}) is encoded in Bayes' Theorem 
    
    \begin{equation}
    \pi\left(\boldsymbol{\theta}|f_{{\rm obs}}\right)\propto\mathscr{L}\left(f_{{\rm obs}}|\boldsymbol{\theta}\right)\cdot\pi_{0}\left(\boldsymbol{\theta}\right),
    \end{equation}
    where $\mathscr{L}$ is the likelihood function, and $\pi_{0}\left(\boldsymbol{\mathbf{\theta}}\right)$ is the prior probability distribution given an array of model parameters $\boldsymbol{\theta}$. Assuming that the likelihood of \cecilia's model parameters can be modelled with a Gaussian distribution, our MCMC is designed to explore and map the multidimensional likelihood function, given by

    \begin{equation}
        \begin{aligned}
        \ln(\mathscr{L}) = &  -\frac{1}{2}\sum_{i=1}^{N_{{\rm points}}}\left[\frac{f_{{\rm obs}}^{i}-f_{{\rm synth,corr}}^{i}}{f_{{\rm obs,err}}^{i}}\right]^{2} + \\
         &+\sum_{j=1}\pi_{0,{\rm phot_{j}}} + \sum_{k=1}\pi_{0,{\rm chondr_{k}}}. 
        \end{aligned}
        \label{eq:mcmc_loglike}
    \end{equation}
    
    Above, we adopt the logarithmic scale for numerical and computational efficiency. In particular, the use of the log-likelihood is driven by three main factors: first, it simplifies and speeds up the calculation of derivatives; second, it penalises unlikely fits by assigning them a very high $\chi^{2}$ value and thus a very small log-likelihood; and third, it does not entail any loss of information during the fit because both $\mathscr{L}$ and $\ln(\mathscr{L})$ have a maximum in the same location.

\begin{table*}
  \centering
  \caption{Abundance accuracy of \cecilia{} as determined from an analysis of 1,000 noiseless synthetic spectra from the test set (see Section \ref{sec:results_synth}). We also show \cecilia's allowed ML bounds as well as the abundance ranges detected in He-rich polluted white dwarfs with effective temperatures between 10,000$\leq$\Teff$\leq$20,000~K (source: default column in the Montreal White Dwarf Database, MWDD; \citealt{Dufour:2016_MWDD}). The dagger symbol ($\dagger$) indicates that \cecilia{}'s performance is stable over the entire ML range of the parameter, while the star symbol ($\star$) is used to highlight that \cecilia{} cannot constrain Be to within 0.2~dex for synthetic data, hence suggesting a potentially worse performance for real, noisy observations. References: [\textit{a}]: \citet{GenestBeaulieu:2019}, [\textit{b}]: \citet{Coutu:2019}, [\textit{c}]: \citet{Wolff:2002}, [\textit{d}]: \citet{Xu:2013},  [\textit{e}]: \citet{Koester:2005}, [\textit{f}]: \citet{Xu:2016}, [\textit{g}]: \citet{Fortin:2020}, [\textit{h}]: \citet{Desharnais:2008}. Last MWDD Access: November 2023. }
  \label{tab:accuracy_table}
  \renewcommand{\arraystretch}{1.4}
  \begin{tabular}{|l|cc|cc|cc|}
    \hline
                & \multicolumn{2}{c|}{Accuracy Bounds} & \multicolumn{2}{c|}{\cecilia{}'s ML Bounds} & \multicolumn{2}{c|}{MWDD}\\ 
                \hline
    Label       &  \textit{$\leq$0.1~dex} & \textit{$\leq$0.2~dex} &  \textit{Min.} & \textit{Max.}   &  \textit{Min.} & \textit{Max.} \\
    \hline \hline  
    \logHHe     &  $\geq$-7$^{\dagger}$   &  $\geq$-7$^{\dagger}$  & -7     & -3    & $<$-6.5$^{a}$  & -2.9$^{e}$     \\  
    \logCaHe    &  $\geq$-11.20           &  $\ll$-12$^{\dagger}$  & -12    & -7    & -10.8$^{b}$    & -6.6$^{f}$    \\  
    \logMgHe    &  $\geq$-7.85            &  $\geq$-9.30           & -16.89 & -0.17 & -8.5$^{c}$     & -5.9$^{g}$     \\  
    \logFeHe    &  $\geq$-8.35            &  $\geq$-11.20          & -18.20 & 0.18  & -8.2$^{c}$     & -5.6$^{g}$    \\  
    \logOHe     &  $\geq$-5.80            &  $\geq$-7.15           & -17.46 & 1.25  & $<$-6.6$^{c}$  & -5.1$^{g}$    \\  
    \logSiHe    &  $\geq$-7.40            &  $\geq$-8.65           & -17.42 & -0.51 & -8.0$^{c}$     & -5.9$^{g}$    \\  
    \logTiHe    &  $\geq$-9.80            &  $\geq$-11.45          & -19.35 & -2.32 & -9.5$^{d}$     & -8.6$^{g}$     \\  
    \logBeHe    & \multicolumn{2}{c|}{$\gg$-5.61$^{\star}$}       & -23.85 & -5.61 &    -     &    -      \\  
    \logCrHe    &  $\geq$-9.20            &  $\geq$-10.60          & -19.61 & -1.71 & -9.3$^{d}$     & $<$-7.7$^{h}$     \\  
    \logMnHe    &  $\geq$-8.25            &  $\geq$-9.80           & -20.08 & -1.65 & -9.8$^{d}$     &  -8.6$^{g}$    \\  
    \logNiHe    &  $\geq$-6.60            &  $\geq$-8.10           & -18.96 & -1.66 & -8.8$^{d}$     &  -7.0$^{g}$   \\  
    \hline
  \end{tabular}
\end{table*}

    In Eq. \ref{eq:mcmc_loglike}, the last two terms represent the prior distributions in our \cecilia{} model. The first term ($\sum_{j=1}\pi_{0, \rm phot_{j}}$) assumes that the user has reasonable estimates for the effective temperature and surface gravity of the white dwarf from an external photometric fit. The second term ($\sum_{k=1}\pi_{0,\rm chondr_{k}}$) is optional and can be used to impose chondritic abundance priors on any metal of interest based on the assumed calcium abundance of the star and the chondritic values of \citet{Asplund:2009}. These physically motivated priors are useful to constrain the abundances of those heavy elements that are barely visible or undetectable in the observed white dwarf spectrum.

    After defining the prior distributions of our likelihood function, we execute our MCMC following the same structure presented in step  \ref{sec:step7_mpfit}; this time, however, we do not minimise the $\chi^{2}$ statistic of \cecilia's predictions, but maximise the log-likelihood function $\ln(\mathcal{L})$. To run our MCMC, we use an ensemble of $n_{\rm walkers}$ walkers and $n_{\rm draws}$ draws, discarding the first $20\%$ of the draws in a ``burn-in'' phase to exclude non-stationary values.\footnote{The walkers and draws represent, respectively, the number of chains in our MCMC, and the number of steps that are evaluated in each chain.} With 1 Satori GPU and suitable combinations of these three MCMC hyperparameters, we find that \cecilia{} can produce an optimal spectroscopic model in less than 10 hours, with the time complexity of this process depending primarily on the the computational time associated to the predictions of the FT FCNN2, rather than the quality and/or resolution of the observed input spectrum. In addition to providing a best-fit spectral solution, our MCMC also determines the conditional probabilities of the model parameters, also known as ``posterior probability distributions.'' From the these posteriors, the MCMC computes the median values of each label, together with their 16th, 50th, and 84th percentiles. In this work, we report our MCMC results using the median statistic and its 1$\sigma$ confidence interval.

    At the end of the MCMC, we also examine the quality of \cecilia's fit with a variety of diagnostic figures and performance metrics. Among them, we generate scatterplot matrices (or ``corner plots'') to visualise the two-dimensional ``marginal'' posterior of a given pair of model parameters alongside their one-dimensional marginal histogram projection. For a parameter $\theta_{i}$, the marginal posterior $\pi\left(\theta_{i}|f_{\rm obs}\right)$ can be computed with
    \begin{equation}
    \pi\left(\theta_{i}|f_{\rm obs}\right) = \int \pi\left(\boldsymbol{\theta}|f_{\rm obs}\right)d\boldsymbol{\theta}_{j\neq i}, 
    \label{eq:marginal_posterior}
    \end{equation}
    where $j$ iterates over all the parameters. Finally, we test the convergence of the chains using the Gelman-Rubin (GR) potential scale reduction factor $\hat{R}$ for each model parameter \citep{GelmanRubin:1992}. This metric compares the within-chain-variance with the scatter of the points between the chains. In this work, we assume convergence of a model parameter if its GR value is lower than 1.003 \citep{Vehtari:2021, Vats:2018}.
\end{enumerate}

\section{Results} \label{sec:results}

In this section, we investigate the performance of \cecilia{} using two types of observations: noiseless synthetic models, and a real spectrum of a well-characterised polluted white dwarf. By testing \cecilia{} under ideal and realistic noisy conditions, we aim to understand its retrieval accuracy, its usage limits, and its suitability for the study of different white dwarfs and astronomical datasets.

\subsection{Sensitivity Analysis with Synthetic Observations} \label{sec:results_synth}

One of the main disadvantages of ML techniques is the lack of their explainability, i.e. the difficulty of interpreting their predictions through a human lens. Despite notable progress in recent years to make ML models more transparent and easier to understand \citep{Saranya:2023, Vilone:2020, Gunning:2019}, neural networks are still perceived as ``black boxes,'' which can sometimes hinder our ability to trust their predictions and their decision-making process. In the context of \cecilia, we have sought to address this problem by evaluating the baseline performance of its trained architecture under well-defined and optimal conditions. 

To conduct our study, we selected 1000 random synthetic spectra from our test set; in other words, we only considered atmosphere models that \cecilia{} had \textit{never} seen during training. We then used \cecilia's FT FCNN2 architecture and the \texttt{mpfit} library to model the spectra. For this task, we fixed \Teff{} and \logg{} to their true values,\footnote{We chose to keep \Teff{} and \logg{} fixed because these parameters are  usually constrained from photometric observations.} and only fitted the 11 elemental abundances listed in Table \ref{tab:ranges_labels} (i.e. from \logHHe{} to \logNiHe). To better understand the retrieval accuracy of \cecilia{} under perfect conditions, we also decided to adopt the true stellar labels of each synthetic spectrum as our input guesses for our optimisation routine. We then executed \texttt{mpfit} with 30 iterations, obtaining a best-fit model in about 1 minute per spectrum. For computational efficency purposes, we did not complement our $\chi^{2}$ minimisation analysis with a full MCMC, and instead treated the \texttt{mpfit}'s results as our final solutions.

\begin{figure*}
    \centering
    \includegraphics[width=0.86\linewidth]{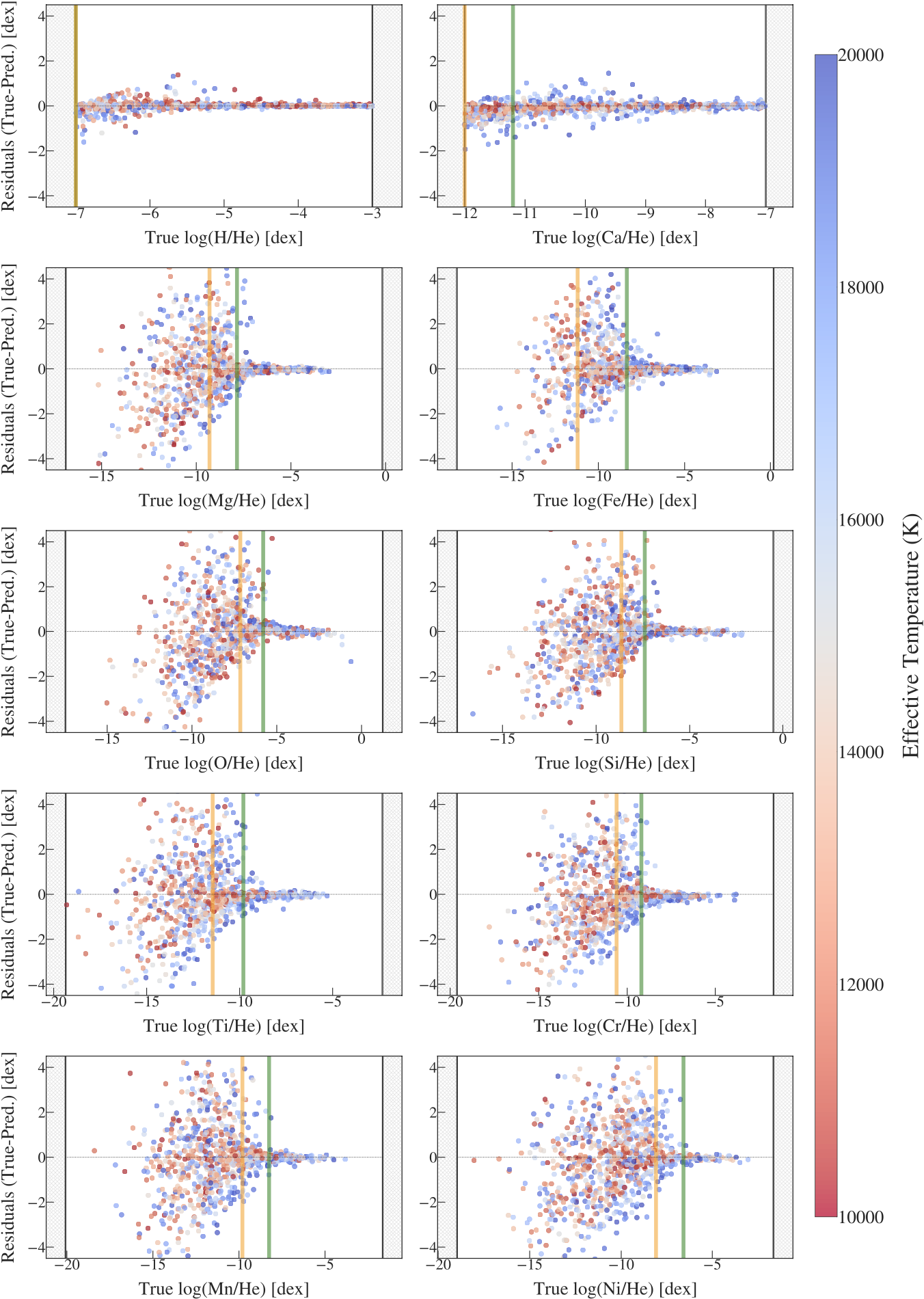}
    \caption{Residuals of \cecilia{} as a function of true  elemental abundance for 1,000 noiseless atmosphere models from our test set (see Section \ref{sec:results_synth}). To conduct our analysis, we fixed \Teff{} and \logg{} to their true values, and only fitted the 11 abundances considered in this work (i.e. from \logHHe{} to \logNiHe{} in Table \ref{tab:ranges_labels}). The vertical lines in green and orange denote, respectively, the boundaries where \cecilia{} achieves a predictive accuracy lower than 0.1~dex and 0.2~dex as calculated by Eq. \ref{eq:accuracy_eqn} (see Table \ref{tab:accuracy_table}). The black lines represent the minimum and maximum ML bounds of \cecilia{} (see Table \ref{tab:ranges_labels}), with the grey hatched regions symbolising abundances outside of \cecilia's allowed parameter space. This figure does not show any results for \Teff{} and \logg{}, which were fixed parameters during our fits. It also does not include our residuals for Be due to the poor performance of our code with this metal. }
    \label{fig:accuracy_residuals_plot}
\end{figure*}

After performing our fits, we determined the residuals $\Vec{r}$ between their true labels and \cecilia's predictions. Our results are presented in \autoref{fig:accuracy_residuals_plot} as a function of photospheric composition, with the colorbar indicating the true effective temperature of the synthetic spectra. Recognising that the visibility of a metal is strongly dependent on \Teff, the general trend observed in \autoref{fig:accuracy_residuals_plot} is that  \cecilia{} performs very well for synthetic white dwarfs with high levels of metal pollution. As metal abundances decrease, the accuracy of our code is also lower because spectral signatures become less strong and harder for \cecilia{} to mimic (typically less than roughly 1$\%$ with respect to the flux continuum, except for a few outliers). In practice, this problem is mitigated by the fact that subtle, shallow lines are challenging to detect observationally as well. 

In \autoref{fig:accuracy_residuals_plot}, we show all the independently varied elements during \cecilia's training except for Beryllium, which our code could not reliably constrain. This light metal has a relatively low abundance in the present-day solar photosphere and in  meteoritic CI chondrites \citep{Asplund:2009, Lodders:2003}, which could also be the case for most polluted white dwarfs. With only two prominent resonance lines of Be in the near ultraviolet at 3,130.42~\angs{} and 3,131.07~\angs{} \citep{NIST_ASD}, its abundance is challenging to determine unless there is a substantial amount of this element in the photosphere of the white dwarf (e.g. \citealt{Klein:2021}). Therefore, given the difficulty of \cecilia{} to model Be, we recommend fixing its abundance to a chondritic ratio before using \cecilia{} to fit the spectrum of a polluted white dwarf. In the remaining sections, we follow this approach to generate our results.

\begin{figure*}
    \centering
    \includegraphics[width=0.9\linewidth]{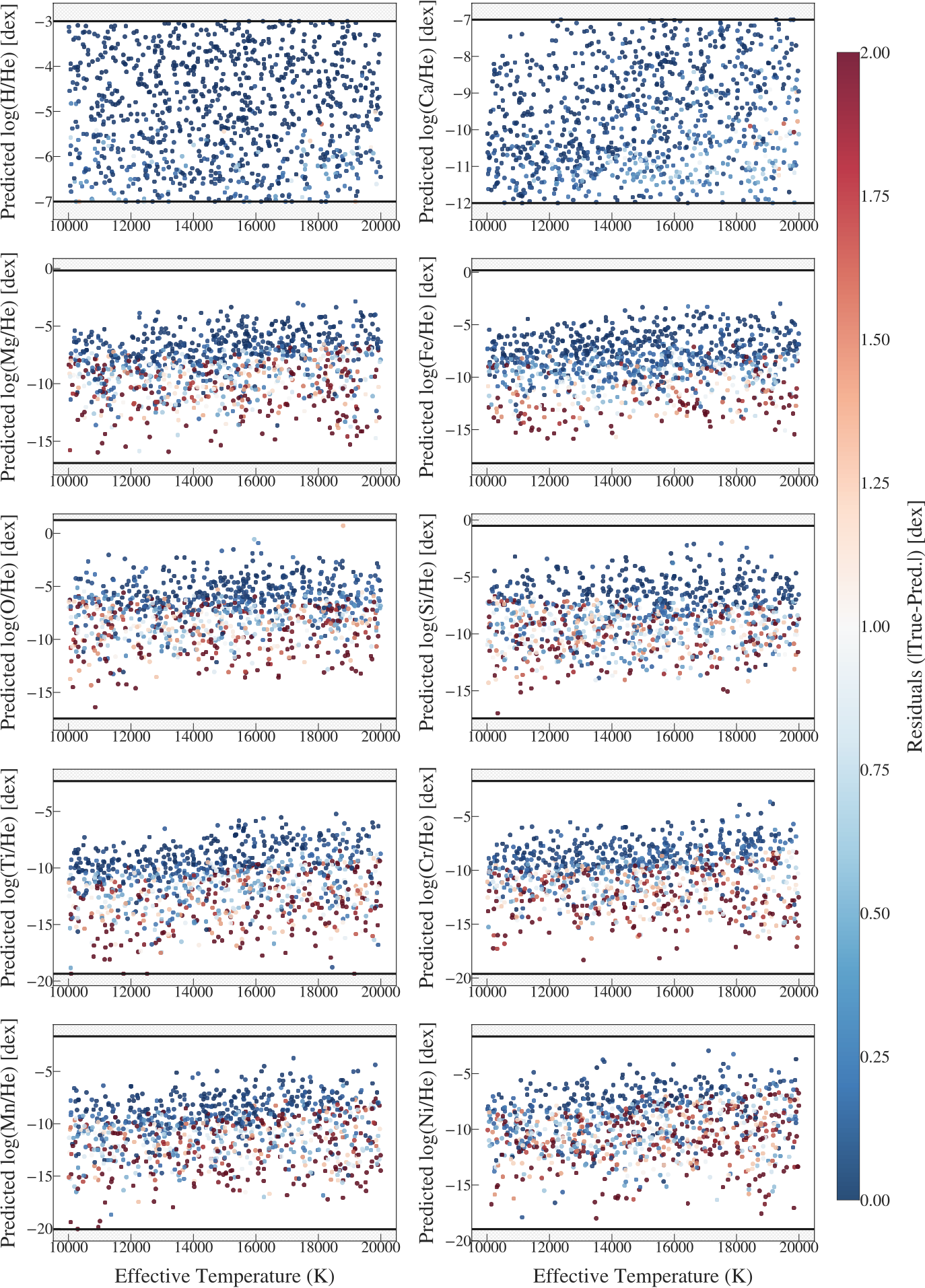}
    \caption{\cecilia's estimated elemental abundances as a function of synthetic effective temperature. The colorbar shows the residuals of its predictions, which are slightly higher for hotter white dwarfs. Similarly to \autoref{fig:accuracy_residuals_plot}, the black lines denote the minimum and maximum allowed ML bounds of Cecilia (see Table \ref{tab:ranges_labels}), while the grey hatched areas represent photospheric compositions outside of \cecilia's range. From this figure, we find that \cecilia's predictions are particularly good for \logHHe{} and \logCaHe.}
    \label{fig:accuracy_teff_plot}
\end{figure*}

Using the residuals from \autoref{fig:accuracy_residuals_plot}, we then calculated the systematic errors associated to \cecilia's \texttt{mpfit} predictions. To this end, we estimated the retrieval accuracy of our pipeline from the ratio between the Median Absolute Deviation (MAD) of the residuals and the median of a standard Gaussian distribution,  
\begin{equation}
    \rm{Accuracy} = \frac{\text{median}\left(|r_{i} - \text{median}(\Vec{r})|\right)}{0.67} \equiv \frac{MAD}{0.67},
    \label{eq:accuracy_eqn}
\end{equation}
where the MAD provides a robust statistical metric against outliers. The green and orange vertical lines in \autoref{fig:accuracy_residuals_plot} show the lowest possible abundance values that \cecilia{} can retrieve with an average accuracy lower than 0.1~dex and 0.2~dex, respectively. These bounds are summarised in Table \ref{tab:accuracy_table} and can be used to assess the usefulness and robustness of \cecilia{} for the study of a real polluted white dwarf. 

\begin{figure*}
  \centering
  \includegraphics[width=0.95\textwidth]{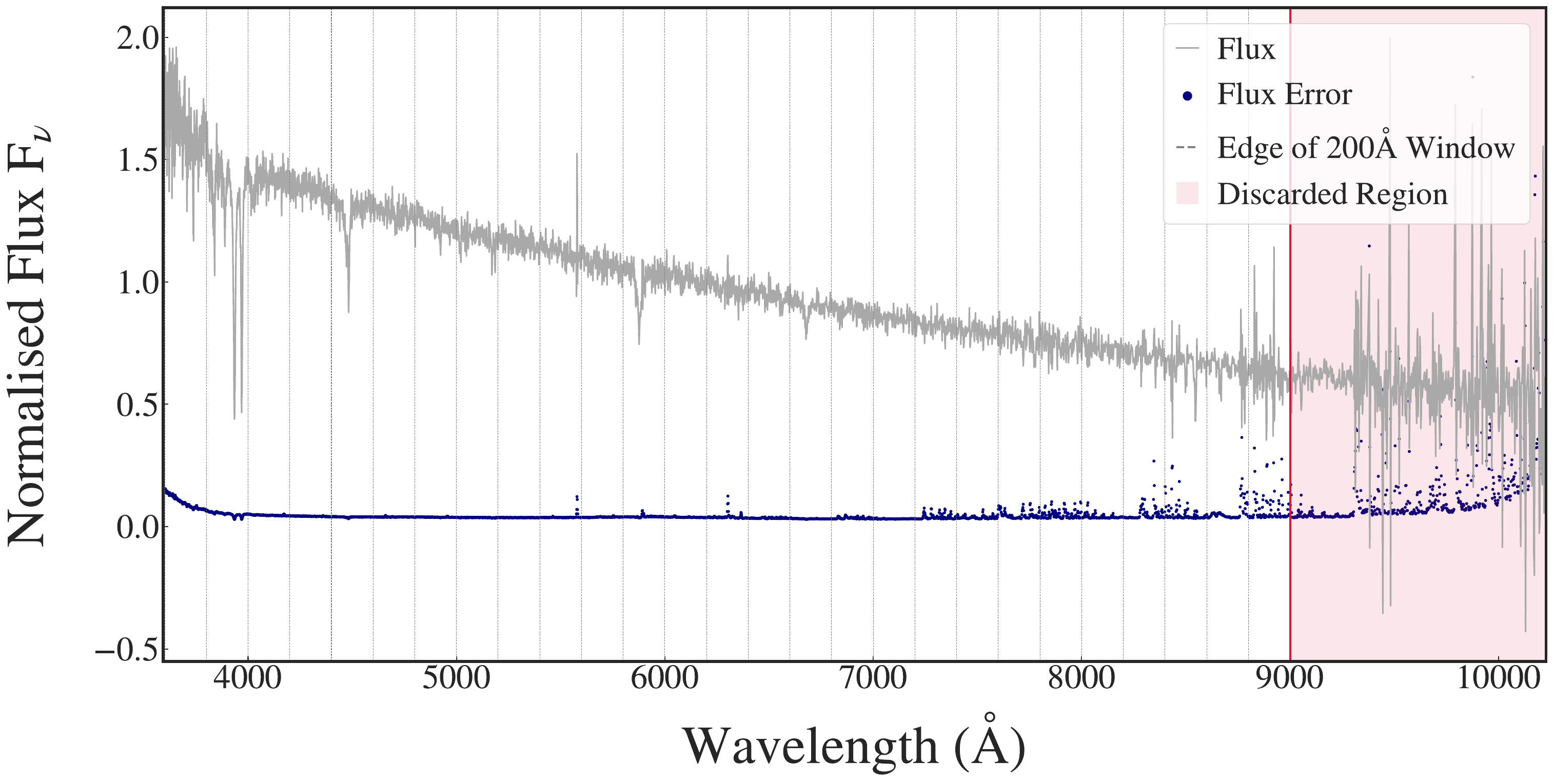}
  \caption{Low-resolution SDSS spectrum of the He-rich, heavily polluted white dwarf \wdreal{}. The normalised $F_{\nu}$ flux and its corresponding observational error are presented in  grey and blue, respectively. The vertical lines denote the spectral windows of 200~\angs{} used to train \cecilia{} (see Section \ref{sec:training}). The red shaded area covers the wavelength region excluded in our optimisation routine (see Section \ref{sec:results_real}).}
  \label{fig:sdss_rawspec_wd1232}
\end{figure*}

\begin{table} 
    \centering
    \caption{Main astrophysical properties of \wdreal. References (Ref.): [1] \citet{Gaia:2016, Gaia:2023}, [2] \citet{York:2000, Almeida:2023}, [3] \citet{Martin:2005_GALEX}, [4] \citet{PanSTARRS:2010}, [5] \citet{Kleinman:2013}, [6] \citet{Coutu:2019}. \label{tab:wd1232_props}}
    \renewcommand{\arraystretch}{1.3}
    \begin{tabular}{|lcc|}
        \hline
        Property & Value & Ref.  \\
        \hline
        \multicolumn{3}{|c|}{\it Other Target Names}  \\
        \hline
        \Gaia{} DR3 &  1571584539980588544  & [1]    \\
        SDSS        &  J123432.65+560643.1  & [2] \\
        \Galex      &  J123432.6+560642     & [3]    \\
        \hline
        \multicolumn{3}{|c|}{\it Astrometric Properties}\\
        \hline 
        R.A. [J2020; h:m:s]	  &  12:34:32.676           & [1] \\
        Dec	[J2020; d:m:s]  & +56:06:43.034             & [1] \\
        Parallax [mas]     & 5.817$\pm$0.095            & [1] \\
        Distance [pc]         & 171.92$\pm$3.00         & [1] \\
        $\mu_{{\rm R.A.}}$ (\masy) & -72.884$\pm$0.089  & [1] \\
        $\mu_{{\rm Dec.}}$ (\masy) & -0.223$\pm$0.094   & [1] \\
        \hline
        \multicolumn{3}{|c|}{\it Photometric Properties}\\
        \hline 
        \Gaia{} G$_{\rm mag}$        & 18.05$\pm$0.01         & [1] \\
        SDSS g$_{\rm mag}$           & 18.27$\pm$0.01	      & [2] \\
        \panstarrs{} g$_{\rm mag}$   & 17.95$\pm$0.01	      & [4] \\
        \hline
        \multicolumn{3}{|c|}{\it Physical and Chemical Properties}\\
        \hline 
        \Teff{}  [K]      &  11,787$\pm$423 &  [6] \\
        \logg{} [cgs]     &  8.30$\pm$0.06  &  [6] \\
        Mass [\Msun]      &  0.77           &  [6] \\
        Cooling Age [Gyr] &  0.64           &  [6] \\
        Atm. Composition  & He              & [5, 6] \\
        Spectral Type     & DBZA            & [5, 6] \\
        \hline
    \end{tabular}
\end{table}

In \autoref{fig:accuracy_teff_plot}, we also illustrate \cecilia's estimated abundances as a function of synthetic effective temperature, with the colorbar representing the residuals of its predictions. From this figure, we find that \cecilia's performance is overall quite stable over the range between 10,000~K and 20,000~K. However, its residuals are slightly higher for high-temperature white dwarfs, which is consistent with the fact that hotter sources are more opaque, thus making it more difficult to detect weak signs of metal pollution \citep[e.g.][]{Zuckerman:2010}. It is also interesting to highlight that \cecilia's predictive power is particularly good for \logHHe{} and \logCaHe{}, as demonstrated by the low error in their predicted abundances. The level of accuracy for these two stellar labels is possibly due to the depth of their absorption lines compared to those of other metals (more than $1\%$ deep relative to the flux continuum, even for the lowest possible abundances of H and Ca at the maximum and minimum allowed values of \Teff{} and \logg{}).


\subsection{Performance with Real Observations}\label{sec:results_real}

In this section, we test and validate \cecilia{} against real spectroscopic observations of the He-rich, heavily polluted white dwarf \wdreal. This star was discovered by \citet{Eisenstein:2006}, who compiled a catalog of 19,712 spectroscopically-confirmed white dwarfs from the Sloan Digital Sky Survey (SDSS, \citealt{York:2000}). More recently, it was studied by \citet{Coutu:2019} and \citet{Xu:2019} at low- and high-resolution: the former conducted a spectroscopic analysis of 1,023 He-atmosphere polluted systems observed by SDSS, while the latter used data from the Keck HIRESb and ESI spectrographs to perform a detailed characterisation of \wdreal{}. From these literature studies, the physical and chemical properties of \wdreal{} appear to fall well within the allowed bounds of \cecilia{} (see Table \ref{tab:wd1232_props}). Therefore, this white dwarf constitutes an excellent candidate to understand the behavior of our pipeline with real astrophysical observations.

To perform our sensitivity analysis, we choose to use \cecilia{} to fit the low-resolution, flux-calibrated spectrum of \wdreal{} from the SDSS DR18 database (plate = 8232, fiber = 84; see \autoref{fig:sdss_rawspec_wd1232}).\footnote{\url{https://skyserver.sdss.org/dr18/}.} Given that SDSS observations have a variable resolving power of $R$=1,500 at 3,800~\angs{} and $R$=2,500~\angs{} at 9,000~\angs{}, we adopt a mean value of \Robs$\approx$2,000 in our fitting procedure. Moreover, to evaluate \cecilia{}'s behavior with unprocessed real data, we do not remove any problematic data points from the spectrum, such as bad flux points or instrumental artefacts. 

As explained in Section \ref{sec:parameter_estimation}, the first step in our fitting routine is to load the SDSS spectrum of \wdreal{} into \cecilia{}, together with reasonable estimates for the white dwarf's effective temperature, surface gravity, elemental abundances (H, Ca, Mg, Fe, O, Si, Ti, Be, Cr, Mn, and Ni), and RV shift. To generate our list of stellar parameters, we adopt the \Teff{} and \logg{} values obtained by \citet{Coutu:2019} from existing photometric observations; the calcium photospheric abundance derived by the same authors from the Ca II H$\&$K absorption lines at air wavelengths 3,934~\angs{} and 3,369~\angs{} \citep{NIST_ASD}; and the elemental abundances of \citet{Xu:2019} for the remaining heavy elements. We also assume a zero RV shift as our initial guess for the RV term in our spectral model. In our analysis, we choose to treat all 14 stellar properties as free model parameters.

\begin{table*}
    \centering
    \caption{MCMC best-fit stellar parameters for the SDSS spectrum of \wdreal{}, together with their uncertainties and assumed ground truths. The systematic ($\sigma_{\rm{sys,\texttt{ML}}}$) and statistical uncertainties ($\sigma_{\rm{stat, MCMC}}$) measure the errors associated to  \cecilia's spectral interpolation (see Section \ref{sec:results_synth}) and to the MCMC fit, respectively. For abundance studies of exoplanetary material, we can approximate the total uncertainty of a model parameter ($\sigma_{\rm{tot}}$) by computing their quadrature sum (see Eq. \ref{eq:total_err}). In this work, if the total error of an abundance is lower than an SDSS noise floor of 0.15~dex, we assume that the MCMC statistical errors are underestimated and impose $\sigma_{\rm{tot}}$$\approx$0.15~dex (see star $\star$ symbol). The dagger symbol ($\dagger$) flags the model parameters dominated by photometric priors (\Teff, \logg). In this table, we do not present the best-fit abundances of the four heavy elements with chondritic priors (Be, Cr, Mn, and Ni); these metals were included in \cecilia's MCMC, but are not detected in the SDSS spectrum. Finally, we do not report the systematic errors for \Teff{} and \logg{} because these two parameters were fixed during our sensitivity study of synthetic spectra (see Section \ref{sec:results_synth}). References (Ref.): [1] \citet{Coutu:2019}, [2] \citet{Xu:2019}.}
    \label{tab:sdss_mcmctab_wd1232}
    \renewcommand{\arraystretch}{1.4}
    \begin{tabular}{|l|cc|cc|cc|}
    \hline 
    &  \multicolumn{2}{c|}{\cecilia's MCMC Best-Fit Model}  & \multicolumn{2}{c|}{Systematic$^{b}$ and Total Uncertainty} & \multicolumn{2}{c|}{Assumed Ground Truth} \\
    \hline
    \textit{Label}$^{a}$ & \textit{Value} & $\sigma_{\rm{stat,MCMC}}$ & $\sigma_{\rm{sys,\texttt{ML}}}$ & $\sigma_{\rm{tot}}$ & \textit{Value} & \textit{Ref.} \\
    \hline\hline 
    \Teff$^{\dagger}$ [K]   & 11777.80 &$^{+86.33}_{-84.81}$      & -       &   $\pm$85.57     & 11787$\pm$423                                 & [1] \\
    \logg$^{\dagger}$ [cgs] & 8.24     & $^{+0.06}_{-0.06}$       & -       &   $\pm$0.06      & 8.30$\pm$0.06                                 & [1] \\
    RV$^{c}$ [km/s]         & 36.09    &$^{+5.41}_{-5.41}$        & $\pm10$ & $\pm$11.37 & 19.0$\pm$2.0      & [2] \\
    \logHHe                 & -5.92    & $^{+0.15}_{-0.19}$      & $\pm0.03$  & $\pm0.17$         & \shortstack{-5.52$\pm$0.10 \\ -5.90$\pm$0.15} & \shortstack{$\rm [1]$ \\ $\rm [2]$} \\
    \logCaHe                & -7.44    & $^{+0.05}_{-0.05}$      & $\pm0.07$  & $\pm0.15^{\star}$ & \shortstack{-7.41$\pm$0.06 \\ -7.69$\pm$0.05} & \shortstack{$\rm [1]$ \\ $\rm [2]$} \\
    \logMgHe                & -6.11    & $^{+0.04}_{-0.04}$      & $\pm0.05$  & $\pm0.15^{\star}$ & -6.09$\pm$0.05                                & [2]       \\
    \logFeHe                & -6.51    & $^{+0.07}_{-0.08}$      & $\pm0.06$  & $\pm0.15^{\star}$ & -6.45$\pm$0.11                                & [2] \\
    \logOHe                 & -5.19    & $^{+0.12}_{-0.13}$      & $\pm0.09$  & $\pm0.15^{\star}$ & -5.14$\pm$0.15                                & [2] \\
    \logSiHe                & -6.45    & $^{+0.10}_{-0.12}$      & $\pm0.07$  & $\pm0.15^{\star}$ & -6.36$\pm$0.13                                & [2] \\
    \logTiHe                & -8.92    & $^{+0.11}_{-0.13}$      & $\pm0.08$  & $\pm0.15^{\star}$ & -8.96$\pm$0.11                                & [2] \\
    \hline
    \end{tabular}
    \vspace{0.1cm}
    \begin{quote}
         \hspace{-0.17cm}\footnotesize{[\it{a}}]: \footnotesize{All abundances are given as logarithmic number abundance ratios relative to Helium, i.e. $\log_{10}$(n(Z)/n(He)).} \\
         \hspace{0cm}\footnotesize{[\it{b}}]: \footnotesize{We note that there is at least another source of systematic errors coming from the atmosphere models themselves; these errors should be of the order of 0.10~dex and are not considered in this work.}\\
        \hspace{0cm}\footnotesize{[\it{c}}]: \footnotesize{We assume a SDSS systematic RV error of 10 km/s based on the estimates provided by e.g. \citet{Abazajian:2009}.}
    \end{quote}
\end{table*}

To begin with, we execute our \texttt{mpfit} least-squares $\chi^{2}$ minimisation routine, restricting the label parameter space to \cecilia's ML bounds, requiring a maximum of 50 iterations, and using gradient step sizes\footnote{In the \texttt{mpfit} code, the ``step size'' corresponds to the step size used to calculate gradients during the fitting process. It does not refer to the size of the parameter update at each iteration of the algorithm for a given model parameter.} of 0.01~K for temperature, 0.01 cgs for surface gravity, 0.01 dex for all the elemental abundances, and 0.0001 km/s for the  RV shift. With these hyperparameters, \texttt{mpfit} generates an optimal spectroscopic solution with its corresponding 14 stellar labels in less than 1.80 minutes. Next, we run a final MCMC with $n_{\rm walkers}$=50 walkers, $n_{\rm draws}$=5,000 draws, and $n_{\rm burn}=0.2\cdot n_{\rm draws}$=1,000 draws. We initialize the walkers in Gaussian balls with standard deviations of 0.01~K for \Teff, 0.01~cgs for \logg, 0.01~dex for the 11 elemental abundances, and 0.0001 km/s for the  RV shift. We then define our likelihood function with Eq. \ref{eq:mcmc_loglike}, incorporating two types of prior distributions: first, we adopt weak photometric priors of 500~K and 0.1 cgs on \Teff{} and \logg{} based on the uncertainties of these parameters in \citet{Xu:2019}; and second, we assume chondritic abundance priors for Be, Cr, Mn, and Ni, which are elements that we cannot visually and confidently detect in the SDSS spectrum.\footnote{We note that we also considered using a chondritic prior on Ti, but we ended up discarding this option after realising that \cecilia{} could constrain this metal relatively well.} For these chondritic labels, we impose a prior width of 0.5 for Be, Cr, and Ni, with a stricter width of 0.25 for Mn to ensure chain convergence.

With the exception of Be, Cr, Mn, and Ni, for which we use chondritic abundance ratios as initial guesses, we take the results of \texttt{mpfit} to initialise our MCMC sampler. This approach, combined with the  \texttt{edmcmc} configuration mentioned above, generates a final spectral model in about 10 hours with a converged Gelman-Rubin statistic lower than 1.008 for all our model parameters. In Table \ref{tab:sdss_mcmctab_wd1232}, we also summarise the MCMC best-fit labels of \wdreal{}, excluding the four undetected heavy elements with chondritic priors (i.e. Be, Cr, Mn, and Ni). We also report the systematic ($\sigma_{\rm sys,\texttt{ML}}$) and statistical errors ($\sigma_{\rm stat,MCMC}$) of each model parameter, obtained respectively from our sensitivity analysis of synthetic spectra (see Section \ref{sec:results_synth}) and our MCMC fit. The former measure the errors of \cecilia's predictions arising from the ML-based spectral interpolation process, while the latter represent the errors of the SDSS data and are hence observational in nature. Assuming Gaussian and uncorrelated errors, we can approximate the total uncertainty of a model parameter ($\sigma_{\rm tot}$) by adding these two sources of errors in quadrature (together with any other systematic errors, or $\sigma_{\rm other}$, coming from the atmosphere models, which we do not consider in this work),
\begin{equation}
    \sigma_{\rm tot} = \sqrt{\sigma_{\rm stat,MCMC}^{2} + \sigma_{\rm sys, \texttt{ML}}^{2} {\textcolor{grey}{(+ \sigma_{\rm other}^{2})}}}.
    \label{eq:total_err}
\end{equation}

Using the above expression, we calculate the total uncertainty of each stellar label and provide it in Table \ref{tab:sdss_mcmctab_wd1232}. As a conservative measure, we assume that the MCMC has underestimated the statistical error of those abundance parameters with $\sigma_{\rm tot}$<0.15~dex, where this threshold represents an approximate SDSS noise floor.\footnote{Our assumed SDSS noise floor of 0.15~dex is a reasonable value based on the authors' experiences with white dwarf spectral modelling.} For these abundances (flagged with a star symbol in Table \ref{tab:sdss_mcmctab_wd1232}), we replace the total uncertainty calculated in this work by $\sigma_{\rm tot}$=0.15~dex.

\begin{figure*}
  \centering
  \includegraphics[width=0.98\linewidth]{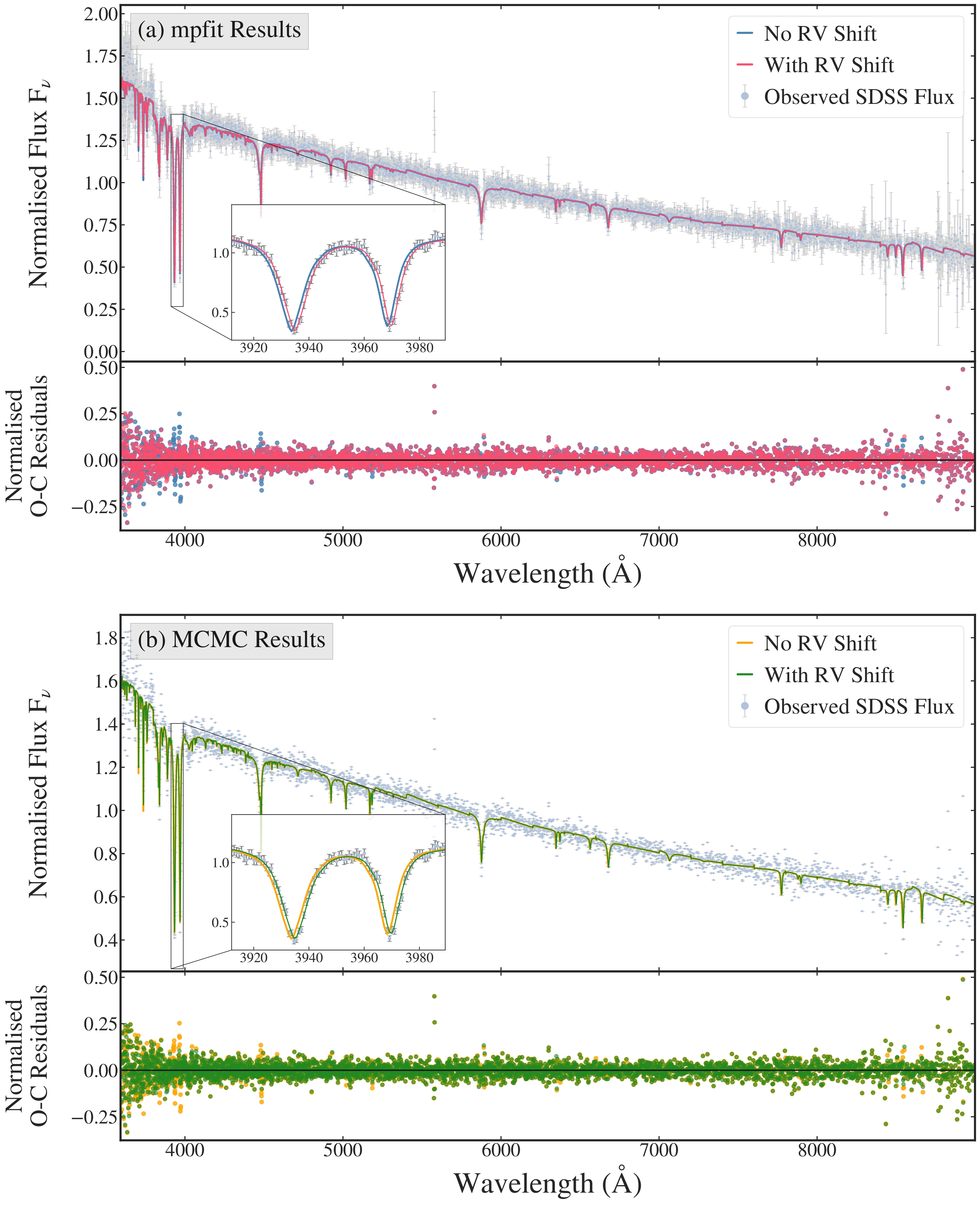}  \caption{Normalised best-fit models  for the SDSS spectrum of \wdreal{} (in light blue). \textit{Panel (a)}: Best-fit \texttt{mpfit} models using an RV shift of 0~km/s and 94.42~km/s (in red and blue, respectively). The inset plot shows the spectral window between 3,800~\angs{} and 4,000~\angs{}, where there are two important Ca II lines at about 3,934~\angs{} and 3,969~\angs{}. The bottom plot shows the normalised flux residuals (Observed-Calculated; O-C) associated to each prediction. \textit{Panel (b)}: Best-fit \texttt{edmcmc} MCMC models using an RV shift of 0~km/s and 94.74 km/s (in green and orange, respectively). Similarly to Panel (\textit{a}), the inset plot shows the spectral window between 3,800~\angs{} and 4,000~\angs{}, while the bottom panel shows the normalised O-C residuals for each spectroscopic solution. We note that the MCMC value of the RV shift is different from that reported in Table \ref{tab:sdss_mcmctab_wd1232} because it also accounts for the barycentric velocity and gravitational redshift correction.}
  \label{fig:sdss_predictions_wd1232}
\end{figure*}

\begin{figure*}
  \centering
  \includegraphics[width=1\linewidth]{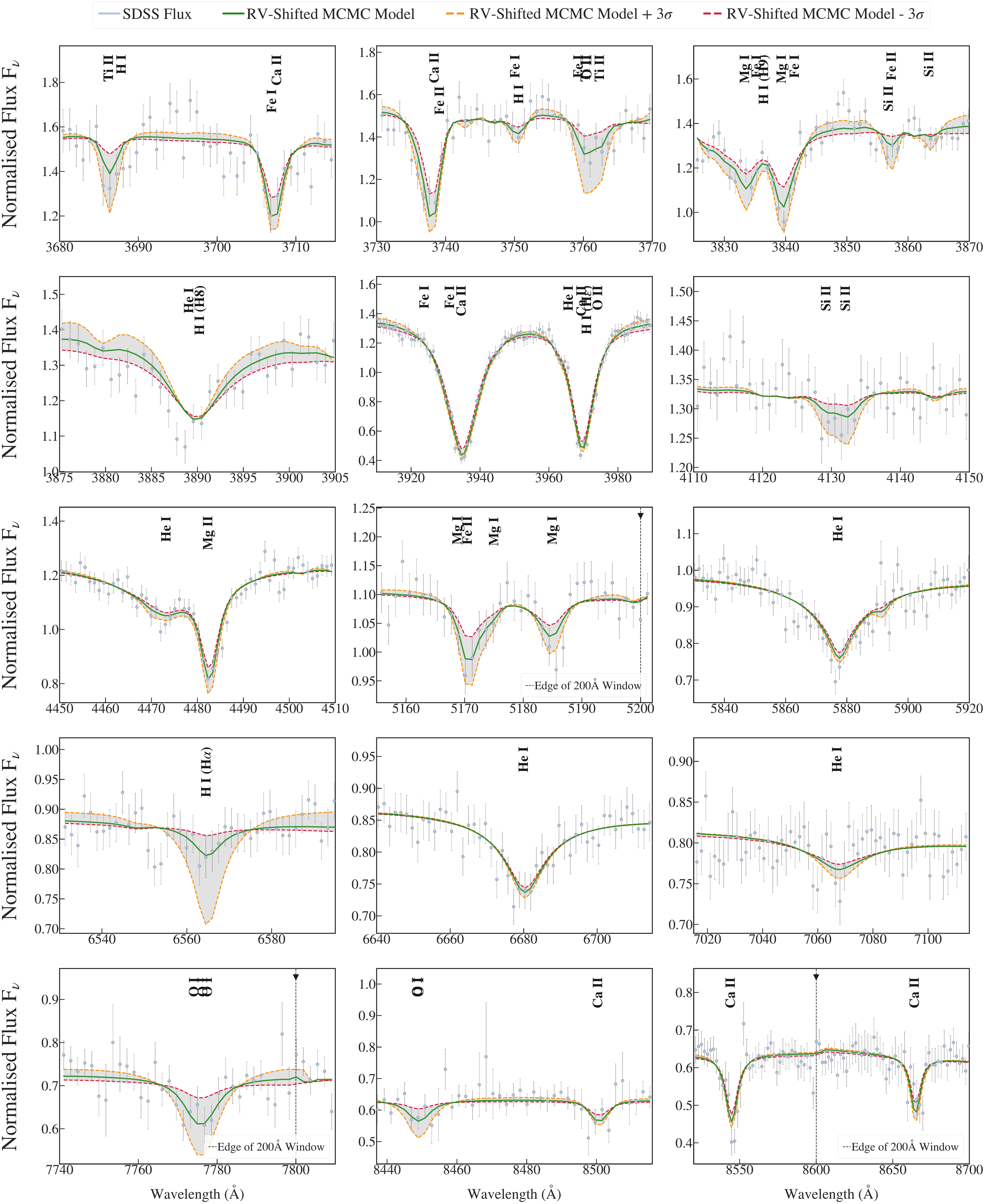}  
  \caption{A selection of spectral windows illustrating our best-fit, RV-shifted MCMC model (in green) for the raw SDSS spectrum of \wdreal{} (in light blue with grey error bars). In each panel, we also show a grey shaded area bounded by an orange and red dashed line, which correspond, respectively, to a $\pm3\sigma_{\rm stat, MCMC}$ variation on the best-fit model (using the same \Teff{} and \logg, and only changing the abundances of the detected elements by $3\sigma_{\rm stat, MCMC}$). In black letters, we indicate some of the strongest spectral signatures of hydrogen, helium, and multiple heavy elements. Finally, we represent the start/end of a 200~\angs{} spectral window with a dashed vertical line and a black inverted triangle symbol.} 
  \label{fig:sdss_mcmcwindows_wd1232}
\end{figure*}

\begin{figure*}
  \centering
  \includegraphics[width=1\linewidth]{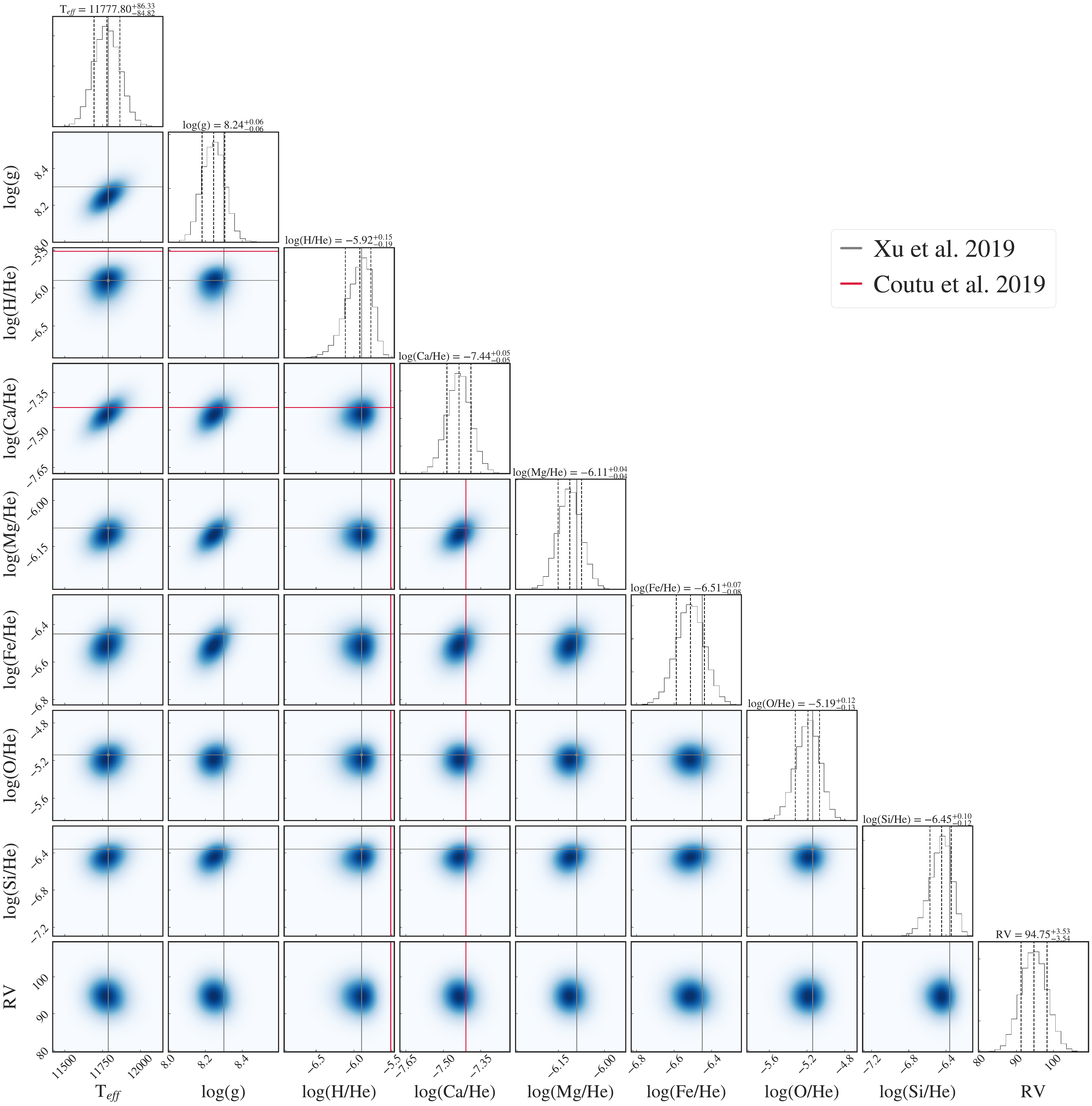}  
  \caption{MCMC corner plot for the SDSS spectrum of \wdreal. The off-diagonal plots show the two-dimensional histograms of the posteriors of the model parameters, marginalised over the rest of parameter values. In contrast, the histograms along the diagonal illustrate their one-dimensional marginalised distributions, together with their corresponding 1$\sigma$ confidence intervals. The grey and red lines show the best-fit values of \citet{Xu:2019} and \citet{Coutu:2019}, respectively, as defined in Table \ref{tab:sdss_mcmctab_wd1232}. In this figure, we exclude the 4 heavy elements with chondritic priors (i.e. Be, Cr, Mn, and Ni) and show the RV shift without subtracting the barycentric motion and gravitational redshift of the white dwarf.}  
  \label{fig:sdss_mcmccorner_wd1232}
\end{figure*}

In \autoref{fig:sdss_predictions_wd1232}, we also show our \texttt{mpfit} and MCMC solutions as well as their corresponding ``Observed-Calculated'' (O-C) flux residuals. To complement this figure, \autoref{fig:sdss_mcmcwindows_wd1232} presents the RV-shifted MCMC model for different regions of the SDSS spectrum with clear metallic absorption lines, together with two additional models obtained with the same \Teff{} and \logg, but with the abundances of the detected elements changed by $3\sigma_{\rm stat,MCMC}$. Finally, \autoref{fig:sdss_mcmccorner_wd1232} presents a corner plot of  the posterior probability distributions of the model parameters as a measure of MCMC chain convergence.

Based on Table \ref{tab:sdss_mcmctab_wd1232}, and taking the properties of \wdreal{} from  \citet{Xu:2019} as our assumed ground truth, our results demonstrate that \cecilia{} can accurately retrieve the majority of parameters in our model. This includes \Teff{} and \logg{} ---for which we use weak photometric priors--- as well as the elemental abundances of 6 heavy elements with visible absorption lines in the SDSS spectrum (Ca, Mg, Fe, O, Si, and Ti; see \autoref{fig:sdss_mcmcwindows_wd1232}). Among these metals, a few of them (e.g. Si, Ti, O) are weak detections, as evidenced by their small observational signatures and their high statistical uncertainties relative to those of other detected elements. For these elements, we might not have been able to claim a confident detection without knowledge of the higher-resolution and signal-to-noise spectra from \citet{Xu:2019}. However, given that our current knowledge of this star confirms their existence, it is promising to see that our pipeline can estimate abundance values consistent with those in the literature. In the future, \cecilia's detection of weak lines in low-resolution data can be a sign that a white dwarf deserves spectroscopic follow-up at high resolution and high signal-to-noise.

\begin{figure}
    \centering
      \includegraphics[width=1\linewidth]{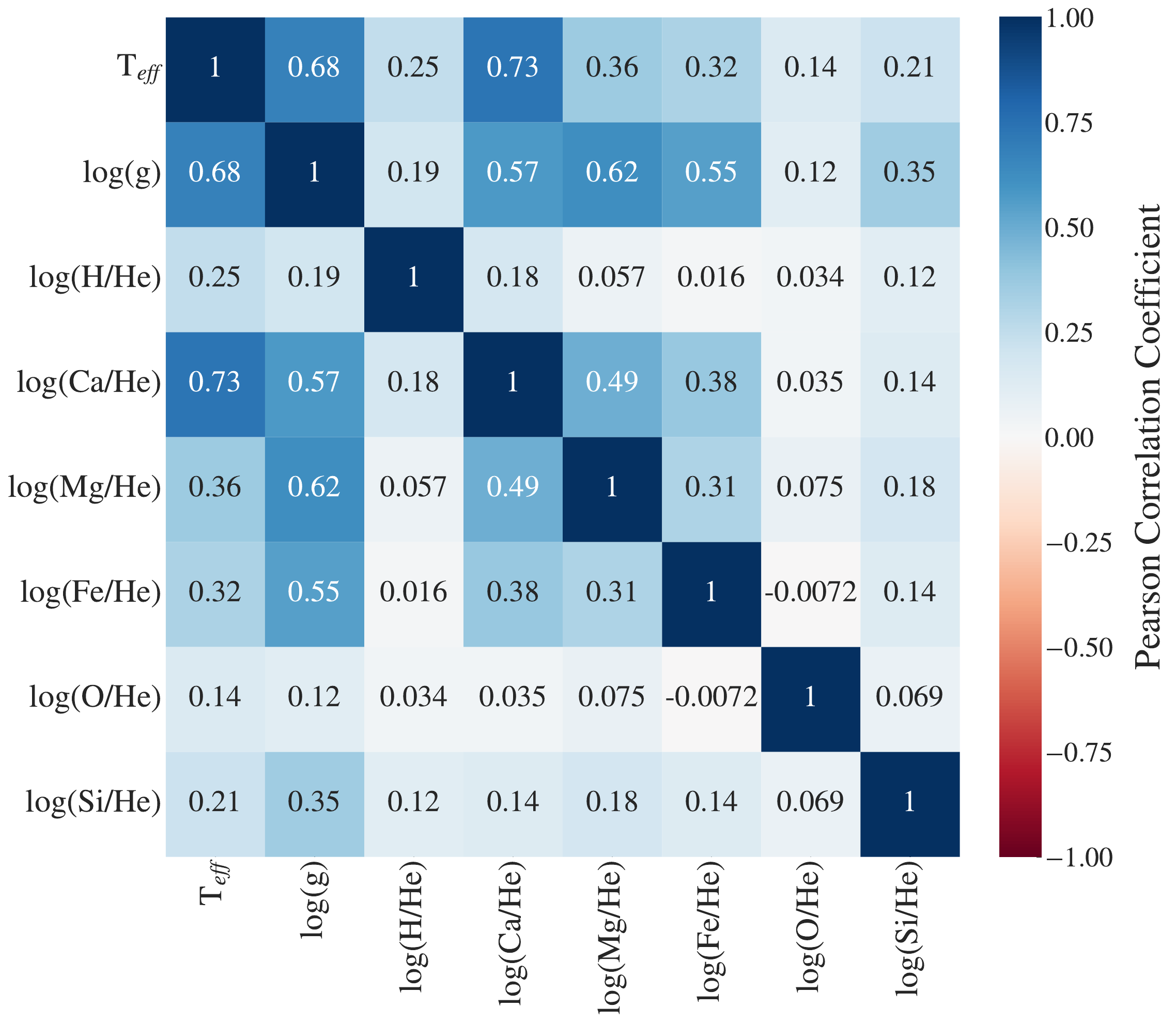}
      \caption{Pearson Correlation Coefficients (PC) for the stellar properties of \wdreal. Some label pairs (e.g. \Teff{} and \logg, \Teff{} and Ca, H and Mg) exhibit a strong positive correlation (PC$\geq$0.5), as suggested by their slanted, oval-like corner plots in \autoref{fig:sdss_mcmccorner_wd1232}. In general, however, most labels have a weak or moderate correlation
      (PC$<$0.5).}
      \label{fig:pearson_coeff}
\end{figure}

With respect to those elements with no clear spectral signatures (Be, Cr, Mn, and Ni), \cecilia{} can marginalise over the uncertainty in the true values of their abundances thanks to the use of chondritic priors in the likelihood function. These chondritic priors limit the values of the parameters explored in our fit, which in turn facilitates MCMC convergence. We note that the assumption of chondritic priors in our MCMC is effectively equivalent to the classical convention of including undetectable metals in atmosphere models with their abundances fixed to a certain value (e.g. chondritic levels).

Another interesting result from Table \ref{tab:sdss_mcmctab_wd1232} is \cecilia's retrieved calcium abundance (\logCaHe$\approx$-7.44), which is more consistent with that of \citet{Coutu:2019} (\logCaHe$_{\rm Coutu}\approx$-7.41) than with the value reported by \citet{Xu:2019} (\logCaHe$_{\rm Xu}\approx$-7.69). This discrepancy is understandable given the different optimisation strategies used in each paper. More specifically, \citet{Coutu:2019} fitted the same SDSS spectrum of \wdreal{} considered in this work, focusing exclusively on the Ca II H$\&$K doublet and using a maximum likelihood approach similar to that of our code. In contrast, \citet{Xu:2019} implemented a line-by-line $\chi^{2}$ minimisation technique of the Ca II H$\&$K doublet and the near-infrared (IR) Ca II triplet in their high-resolution Keck spectrum. For this task, they individually fitted all the Ca lines and weighted them equally to obtain a mean calcium abundance, hence giving more importance to the IR triplet than \cecilia{} does. Indeed, our code fits all the Ca lines simultaneously ---in combination with the full spectrum--- and subsequently weighs them based on their observed strengths, uncertainties, and signal-to-noise. As \autoref{fig:sdss_mcmcwindows_wd1232} shows, the Ca II H$\&$K lines in the SDSS data are significantly stronger than the near-IR Ca II triplet at around 8,498~\angs, 8,542~\angs and 8,662~\angs{}. Therefore, similarly to the study of \citet{Coutu:2019}, they dominate \cecilia's fitting procedure, leading to a best-fit \logCaHe{} abundance somewhat higher than the result of \citet{Xu:2019}. In addition to the different fitting methodologies employed in each paper, it is also possible that the uncertainties on the SDSS flux, which are higher in the red region of the spectrum (likely due to telluric and/or instrumental effects), may have brought our Ca abundance closer to that of \citep{Coutu:2019} than to the result of \citep{Xu:2019}. 

Besides providing the main physico-chemical properties of \wdreal{}, \cecilia{} generates the posterior probability distributions of its model parameters and shows their interdependence in the form of a corner plot (see \autoref{fig:sdss_mcmccorner_wd1232}). With this Bayesian approach, \cecilia{} differs from conventional white dwarf analysis techniques, which can only use a limited number of points in their interpolation grid to explore the relationship between two model parameters \citep[e.g.][]{Klein:2011, Xu:2019}. In general terms, corner plots reveal possible correlations between two labels if their posterior distributions have a slanted and oval-like contour shape. For \wdreal{}, this seems to be the case for multiple stellar properties, such as \Teff{} and \logg{}, \Teff{} and \logCaHe, \logg{} and \logMgHe, or \logCaHe{} and \logMgHe. To qualitatively investigate the presence of these correlations, we computed the Pearson Coefficient (PC) for all our model parameters,
\begin{equation}
    \mathrm{PC}=\frac{\sum (\pi_{x} - \mu_{\pi_{x}})(\pi_{y} - \mu_{\pi_{y}})}
         {\sqrt{\sum (\pi_{x} - \mu_{\pi_{x}})^2 \sum (\pi_{y} - \mu_{\pi_{y}})^2}}\subset[-1,1],
         \label{eq:pearson}
\end{equation}
where $\pi_{x}$ and $\pi_{y}$ are the posterior distributions of labels $x$ and $y$, and $\mu_{\pi_{x}}$ and $\mu_{\pi_{y}}$ are the arithmetic means of their posteriors. A PC greater or lower than 0 denotes, respectively, a positive or negative correlation, while a PC=$\pm1$ indicates that the data can be perfectly modelled by a line.

In \autoref{fig:pearson_coeff}, we present the Pearson coefficients for \cecilia's best-fit MCMC model parameters. Assuming a strong positive correlation if PC$\geq$0.5, our results suggest that certain stellar labels are linearly dependent on one another (e.g. \Teff{} and \logg, or \logCaHe{} and \logMgHe), with an increase in one label resulting in an increase of its corresponding label pair. Based on the corner plot and the Pearson coefficients alone, it is hard to determine whether the observed correlations can be generalised to the entire population of polluted white dwarfs. Although such an investigation is outside the scope of this paper, it is possible that certain stellar properties are linearly dependent on one another due to observational or physical reasons. Observationally, some of the  metal correlations may be caused by the presence of blended absorption lines. From a physical perspective, changes in \Teff{} and \logg{} (a proxy for pressure) modify the temperature-pressure conditions at the stellar photosphere, which in turn control the strength and the width of the absorption lines. In addition, different values of \Teff{} and pressure result in different excitation, ionisation, or dissociation states of the atoms in the stellar photosphere (e.g. Saha equation, Boltzmann equation), which could also potentially affect the metal abundances of the white dwarf.

\section{Discussion}\label{sec:discussion}

\subsection{Implications for White Dwarf Science} \label{sec:implications_wd_science}


As demonstrated in Section \ref{sec:results},  \cecilia{} is the first pipeline capable of simultaneously estimating 10 elemental abundances with a retrieval accuracy of $\lesssim$0.1~dex. This performance compares favorably to that of conventional techniques, which in turn depend on a variety of factors, such as the main properties of the star (e.g. \Teff, \logg, level of metal contamination), the quality of the spectroscopic observations (e.g. resolution, signal-to-noise), or the constitutive physics of their model atmosphere codes. In particular, conventional methods can reach typical uncertainties of about 0.10-0.20~dex for He-rich polluted white dwarfs, both across the UV and the optical range, and irrespective of their underlying atmosphere models \citep[e.g.][]{Doyle:2023, Klein:2021, Izquierdo:2020, Raddi:2015, Wilson:2015, Jura:2012, Zuckerman:2007}. Therefore, \cecilia{}'s performance is close to ---if not comparable to--- that of classical approaches, only introducing small uncertainties due to the accumulation of errors in its ML predictions. With a promising performance and an automated architecture that eliminates the need for manual, time-consuming, and iterative work, \cecilia{} stands as a promising tool for the study of polluted white dwafs and the bulk composition of their accreted  material. 

With the advent of wide-field, multi-object astronomical surveys like SDSS-V \citep{Kollmeier:2017_SDSSV}, the Dark Energy Spectroscopic Instrument (DESI; \citealt{DESIa_Cooper2023, DESIb}), or WEAVE \citep{Dalton:2014_WEAVE}, the volume of spectroscopic observations is poised to grow exponentially \citep{Smith:2023}. These surveys are expected to acquire a large number of white dwarf spectra, so they represent a unique opportunity to improve our fundamental knowledge of metal pollution. Unfortunately, as of today, the exploitation of these datasets remains a challenge for conventional techniques due to their time-consuming and human-supervised nature. With \cecilia{}, we have developed a practical solution to characterise the spectra of polluted white dwarfs in a fast, automated, and accurate way. For instance, using 1 GPU of the MIT Satori supercomputer and limiting our optimisation procedure to  \texttt{mpfit}, we estimate that \cecilia{} would take about 14 and 17 days to process the 40,000 and 50,000 white dwarf spectra in the cumulative DESI and WEAVE databases, respectively. In reality, \cecilia{}'s performance should be much faster than two weeks, as only a small fraction of these observations will exhibit signatures of metal pollution. To quickly identify these polluted spectra, \cecilia{} could benefit from other efficient ML-based classificators, such as the deep neural networks of \citet{Vincent:2023} or the Random Forest approach of \citet{GarciaZamora:2023}.

Another valuable point of comparison is the SDSS-V survey, which has monitored more 6,000 unique \Gaia{} white dwarfs as of 2021, and will acquire a total of 100,000 spectra of these stars in the coming years \citep{Kollmeier:2017_SDSSV, Chandra:2021}. Assuming, unrealistically, that all these spectra will be metal-polluted, they would represent an intractable amount of data for traditional white dwarf characterisation methods. Nonetheless, with \texttt{mpfit} only, it would only take about a month to process all these observations. Therefore, in light of these capabilities, our ML-based pipeline is a powerful tool to harness the information content of massive spectroscopic databases and scale up individual studies of polluted white dwarfs, thus opening the door to a statistically-informed understanding of ancient extrasolar geochemistry. Ultimately, we aspire to use \cecilia\ to bring the field of white dwarf science into the era of big data, allowing scientists to overcome the scalability problem associated to ``human-in-the-loop''  conventional analysis techniques.

In addition to \cecilia's promising retrieval accuracy and speed, our pipeline offers a innovative Bayesisan framework to estimate the elemental abundances of polluted white dwarfs. More specifically, our inference approach differs from that of classical methods because (i) it integrates prior knowledge about the spectra and the main properties of the white dwarf, (ii) it provides robust parameter uncertainties based on their full posterior probability distributions, and (iii) it illustrates possible degeneracies between different model parameters thanks to the use of corner plots. Through these added capabilities, \cecilia{} aims to provide a more detailed and nuanced perspective on the properties of white dwarf pollutants and their parent bodies.  

\subsection{Limitations and Future Work}

The sensitivity analyses presented in Section \ref{sec:results} show that \cecilia{} can effectively determine the main properties of He-rich polluted white dwarfs, achieving an accuracy comparable to that of conventional techniques without the need for human supervision. This  performance has the potential to unlock population-wide studies of metal pollution, but it has several limitations that deserve further investigation. In this Section, we highlight the main caveats of our pipeline, discuss possible mitigation strategies, and propose new opportunities for future work. 

\subsubsection{General ML Design Choices}

First, \cecilia's ML architecture can be improved along three important dimensions: versatility, speed, and accuracy. In terms of versatility, our work has revolved around He-dominated polluted white dwarfs, which ---for a given abundance level--- exhibit stronger metallic lines than H-rich stars due to their lower photospheric opacity \citep{Dufour:2012, Klein:2021, Saumon:2022}. Nevertheless, \cecilia's potential extends beyond He-dominated systems, as its ML architecture can always be retrained with the model atmospheres and astrophysical properties of other stellar objects. 

With respect to speed, our full optimisation procedure takes between 4 and 10 hours to provide a best-fit spectroscopic solution, depending on the user's choice of MCMC hyperparameters (see Section \ref{sec:parameter_estimation}). This behavior makes \cecilia{} an efficient pipeline capable of modelling several stars at the same time, while running uninterruptedly on a high-performance supercomputer cluster. The scalability of \cecilia{} represents an improvement over the performance of classical methods, but it can still be optimised further in preparation for large-scale studies of polluted white dwarfs. A possible solution to making \cecilia{} more efficient would be to explore whether simpler and/or faster ML architectures can achieve similar performance.

Lastly, \cecilia's retrieval accuracy is better than 0.1~dex for 10 heavy elements, which places our pipeline at the level of many conventional techniques. Nevertheless, the errors accumulated in \cecilia's networks introduce a small amount of noise into its final synthetic prediction, which in turn challenges the modeling of very shallow lines and introduces limits below which \cecilia{} performs poorly. As model atmospheres improve with our understanding of white dwarf physics, larger training sets and/or better ML architectures might help improve \cecilia's predictive abilities, obtaining a better match between its best-fit solutions and the observed metal abundances of a polluted white dwarf. Regardless of whether we improve \cecilia's ML predictions, the errors are small enough that most absorption lines detected in the data would be well modelled by our code.

\subsubsection{Training Improvements}

As explained in Section \ref{sec:data}, the size and quality of a training dataset can highly affect the performance of a neural network. In this work, we have have trained \cecilia{} with more than 22,000 synthetic labels and atmosphere models, assuming that white dwarfs can be characterised with a set of 13 randomly varied independent parameters (\Teff, \logg, \logHHe, and 10 metal abundances). As we design future iterations of \cecilia, a larger and more comprehensive training dataset might help our pipeline to achieve a better level of precision. Another important update of \cecilia{} would involve changing our training strategy to prevent distortions of the lines near the edges of the 200~\angs{} spectral windows. A possible solution to this problem would be to re-train \cecilia{} using a system of overlapping windows.

In relation to the stellar labels used in this work, we have identified three improvement opportunities that can be easily implemented in the future. To begin with, we can extend the abundance ranges listed in Table \ref{tab:ranges_labels} ---upwards and downwards--- so that \cecilia{} can gain sensitivity to both stronger and weaker levels of metal pollution. This is particularly relevant for Ca and Be. For instance, the inclusion of higher Ca abundances in our training set would allow us to validate our code against several heavily polluted white dwarfs whose astrophysical properties are currently outside of \cecilia's allowed ML bounds (e.g. GD 40, Ton 345, GD 362, although the latter would also require extending \cecilia{} to lower temperatures, as discussed below). Similarly, expanding the abundance ranges of Be to higher values would make our pipeline more sensitive to stronger Be lines. This would facilitate the analysis of several Be-polluted systems such as those described in \citet{Klein:2021}, which have a $\log(\rm Be/Ca)$ ratio ($\approx$-2.50~dex) considerably higher than that of the Sun ($\approx$-4.96~dex, \citealt{Asplund:2009}). Second, we can make our pipeline more useful by retraining it  with more heavy elements. With this idea in mind, we aim to include at least three new elements: carbon (C), sodium (Na), and aluminium (Al). These species have been detected in the atmospheres of multiple white dwarfs and can provide valuable insights into the volatile (C) and rocky (Na, Al) compositions of white dwarf pollutants (\citealt{Greenstein:1976, Holberg:1998}; see review by \citealt{Klein:2021}). Finally, we can try to improve \cecilia's performance by exploring different ways to generate our stellar labels. This is especially important for the 14 metals with fixed chondritic abundances (C, N, Li, Na, Al, P, S, Cl, Ar, K, Sc, V, Co, and Cu), which might have introduced biases in our pipeline and caused \cecilia's spectral predictions to favor chondritic values. Although the results presented in Section \ref{sec:results} do not show any evidence of this behavior, it will be important to investigate this potential problem in future versions of \cecilia{}. 

Regarding our atmosphere models, there are two possible changes that might make \cecilia{} more useful and versatile. On the one hand, we can retrain our pipeline using a broader wavelength coverage. At present, \cecilia{} can analyze spectroscopic observations in the optical range between 3,000~\angs{} and 9,000~\angs{}, but it is not prepared to model data in the UV ($\sim$900$\leq \lambda\leq$3,000~\angs{}). The UV contains important spectral signatures of volatile species ---e.g. oxygen, carbon, nitrogen, sulfur, phosphorous--- and is thus a crucial region for understanding the volatile budget of a white dwarf pollutant. From a practical point of view, the integration of the UV into our pipeline will also make \cecilia{} particularly relevant for the study of UV observations from space-based facilities such as the Hubble Space Telescope. 

Moreover, we can also retrain \cecilia{} with synthetic spectra covering a broader range of effective temperatures. For example, it could be useful to include models of slightly cooler (\Teff$<$10,000~K) and hotter white dwarfs (20,000$\leq$\Teff$\leq$25,000~K). At temperatures higher than \Teff$>$25,000~K, the origin of metal pollution is unclear due to the outward effects of radiative levitation pressure, which can bring metals to the photosphere from the deep stellar interior \citep{Chayer:1995}. For cooler white dwarfs, it is still possible to observe traces of metal pollution in their spectra \citep[e.g.][]{Elms:2022, Blouin:2020, Hollands:2017}. While most of these cool systems only show a handful of metallic absorption lines (because of the insufficient thermal energy to populate atomic states responsible for many transitions in the optical), they also allow the detection of elements that cannot be easily observed in warmer objects where they are ionized (e.g. Na, K, Li).

\subsection{Fitting Analysis}

In addition to exploring possible improvements to \cecilia's ML architecture and training datasets, we have also identified four opportunities to refine our fitting procedure. First, we can reconfigure our code to predict upper abundance limits when the absorption features of a metal are ambiguous or unobservable. This option is currently unavailable because our pipeline was trained in logarithmic space, which implies that \cecilia{} can never predict a zero elemental abundance (i.e. the complete lack of an element). Introducing the possibility of quantifying upper limits would involve a reformulation of our code, but would constitute a valuable addition to \cecilia. 

A second enhancement to our pipeline would be to allow for a simultaneous spectroscopic and photometric fit of a polluted white dwarf spectrum. Given that our code does not currently use photometric observations in its optimisation routine, \cecilia{} requires the user to have accurate estimates of the effective temperature and surface gravity of the star. These two parameters are also dependent on the amount of heavy elements in the stellar photosphere, so their prediction becomes an iterative problem.\footnote{We note that this issue also affects traditional white dwarf characterisation methods. As of today, several iterative approaches have been developed to mitigate it \citep[e.g.][]{Dufour:2005, Dufour:2007, Coutu:2019}. These approaches perform an initial fit to the stellar photometry, followed by a spectroscopic fit. Afterwards, this optimisation process is repeated multiple times ---each time with a better set of initial parameters--- until a best-fit solution is found \textit{both} for the existing photometric and spectroscopic observations.} Moreover, even if these two properties are well-defined, they might have been estimated with a set of atmosphere models different to those employed in this work, which could potentially introduce minor errors in \cecilia's predictions. Therefore, to ensure that our parameter estimation routine is entirely self-consistent, it would  be beneficial to incorporate a simple neural network capable of estimating values for \Teff{} and \logg{} based on the metal abundances predicted by \cecilia{} at each iteration. 

Third, we hope to improve \cecilia's MCMC fitting procedure by introducing a jitter term in our log-likelihood function. This parameter would allow us to fit unknown sources of noise (e.g. instrumental noise, stellar activity), and could thus be particularly important to model weak absorption lines in real spectroscopic observations. Finally, given that MCMC pipelines can sometimes be quite sensitive to the initialisation of the model parameters, we hope to make \cecilia{} capable of calculating the RV shift of the white dwarf when no estimates are provided by the user.

\section{Conclusions}\label{sec:conclusions}

In this work, we have presented \cecilia, the first ML pipeline designed to measure the main physical and chemical properties of intermediate-temperature, He-rich polluted white dwarfs from their spectra. In particular, our code exploits the power of neural-network-based-interpolation to (i) rapidly generate synthetic spectral models of polluted white dwarfs in high-dimensional space, (ii) map the latter to a set of 13 underlying stellar properties (\Teff, \logg, \logHHe, and 10 metal abundances); and (iii) generate a final spectral prediction through frequentist and Bayesian fitting techniques (see Sections \ref{sec:data} and \ref{sec:methods}). 

With the development of \cecilia, we have sought to tackle the dependence of conventional spectral analysis techniques on manual and iterative work, which in turn make them time-intensive, prone to human errors, and impossible to scale up to large samples of polluted white dwarfs. As explained in Section \ref{sec:results}, \cecilia's architecture speeds up and automates these conventional methods by providing preliminary metal abundances in less than 2 minutes and a final spectroscopic solution in less than 10 hours. After testing the performance of our pipeline with noiseless synthetic data and with real spectroscopic observations of \wdreal, we have shown that \cecilia{} can retrieve metal abundances with uncertainties better than 0.1~dex for up to 10 elements, with only beryllium proving hard to constrain. This level of accuracy is similar to that of state-of-the-art techniques, and it can only be improved in the future as we incorporate larger/better training datasets and integrate user feedback. 

Acknowledging the main limitations of our pipeline (see Section \ref{sec:discussion}), \cecilia{} has the potential to enable population-wide studies of metal pollution without entailing any manual work. These studies will significantly expand the number of polluted white dwarfs with well-characterised abundances, thus playing an important role in improving our knowledge of the geology and chemistry of extrasolar material. Unveiling the diversity of extrasolar compositions is especially relevant in the context of upcoming wide-field astronomical surveys like DESI, WEAVE, or SDSS-V, which will acquire a large volume of spectroscopic observations in the coming years. While conventional white dwarf characterisation techniques are too slow to process these massive datasets, our ML pipeline offers a fast, automated, and reliable solution to measuring trace elements in the atmospheres of polluted white dwarfs. 

Finally, as a feature extraction tool with a versatile ML architecture, \cecilia{} can also be replicated and transferred to other disciplines beyond the realm of white dwarf and of planetary science. As we step into the world of Big Data, where the ability to uncover meaningful insights from observations will increasingly become more important, \cecilia{} exemplifies the immense potential of combining powerful ML tools with state-cutting-edge scientific knowledge and analysis techniques.

\section{Data Availability}

The SDSS spectrum of \wdreal{} can be downloaded from the SDSS DR18 online database.



\section*{Acknowledgements}

MBA thanks Vedant Chandra, Oriol Abril-Pla, Dr. Kishalay De, Dr. Ignasi Ribas, and Dr. Amy Bonsor for useful discussions. MBA is supported by the MIT Department of the Earth, Atmospheric, and Planetary Sciences, NASA grants 80NSSC22K1067 and 80NSSC22K0848, and the MIT William Asbjornsen Albert Memorial Fellowship. SB is a Banting Postdoctoral Fellow and a CITA National Fellow, supported by the Natural Sciences and Engineering Research Council of Canada (NSERC). SX is supported by NOIRLab, which is managed by the Association of Universities for Research in Astronomy (AURA) under a cooperative agreement with the National Science Foundation.

This work has made use of the Montreal White Dwarf Database \citep{Dufour:2016_MWDD}.

This work has made use of data from the European Space Agency (ESA) mission \Gaia{} (\url{https://www.cosmos.esa.int/gaia}), processed by the \Gaia{} Data Processing and Analysis Consortium (DPAC,
\url{https://www.cosmos.esa.int/web/gaia/dpac/consortium}). Funding for the DPAC has been provided by national institutions, in particular the institutions participating in the \Gaia{} Multilateral Agreement. 

This work has also made use of data from the Sloan Digital Sky Survey (SDSS), which is funded by the Alfred P. Sloan Foundation, the Heising-Simons Foundation, the National Science Foundation, and the Participating Institutions. SDSS acknowledges support and resources from the Center for High-Performance Computing at the University of Utah. The SDSS web site is \url{www.sdss.org}. SDSS is managed by the Astrophysical Research Consortium for the Participating Institutions of the SDSS Collaboration, including the Carnegie Institution for Science, Chilean National Time Allocation Committee (CNTAC) ratified researchers, the Gotham Participation Group, Harvard University, Heidelberg University, The Johns Hopkins University, L'\'Ecole polytechnique F\'ed\'erale de Lausanne (EPFL), Leibniz-Institut fur Astrophysik Potsdam (AIP), Max-Planck-Institut fur Astronomie (MPIA Heidelberg), Max-Planck-Institut fur Extraterrestrische Physik (MPE), Nanjing University, National Astronomical Observatories of China (NAOC), New Mexico State University, The Ohio State University, Pennsylvania State University, Smithsonian Astrophysical Observatory, Space Telescope Science Institute (STScI), the Stellar Astrophysics Participation Group, Universidad Nacional Aut\'onoma de Mexico, University of Arizona, University of Colorado Boulder, University of Illinois at Urbana-Champaign, University of Toronto, University of Utah, University of Virginia, Yale University, and Yunnan University.

This work has employed the following open-source software packages: \texttt{Python} \citep{Python}, \texttt{numpy} \citep{numpy}, \texttt{scipy} \citep{scipy}, \linebreak 
\texttt{matplotlib} \citep{matplotlib}, \texttt{pandas} \citep{pandas:2010}, \texttt{mpfit} \citep{Markwardt:2009}, \texttt{edmcmc} \citep{Vanderburg:2021_edmcmc}, \texttt{tensorflow} \citep{tensorflow2015-whitepaper}, and \texttt{corner} \citep{corner, corner_luger}.



\bibliographystyle{mnras}
\bibliography{ref} 








\bsp	
\label{lastpage}
\end{document}